\documentclass{aa}  

\usepackage{graphicx}
\usepackage{txfonts,textcomp}
\usepackage{hyperref}

\hypersetup{colorlinks=true,linkcolor=blue,citecolor=blue,filecolor=blue,urlcolor=blue}
\newcommand{\s}{\mbox{SRGt 062340}}

\newcommand{\xmm}{\textit{XMM-Newton}}
\newcommand{\ero}{eROSITA}
\begin{document} 

   \title{Confirmation of \mbox{SRGt 062340.2-265751} as a nova-like cataclysmic variable with a possible magnetic nature}

   \titlerunning{On the nature of SRGt 062340.2-265751} 
   \authorrunning{V.~A.~C\'uneo et al.}

   \author{V.~A.~C\'uneo\thanks{Corresponding author: \texttt{virginiacuneo@gmail.com}}\inst{1} \and
   A.~D.~Schwope\inst{1} \and
   J.~Kurpas\inst{1} \and
   A.~Avakyan\inst{2} \and
   J.~Brink\inst{1,3} \and 
   D.~A.~H.~Buckley\inst{4,5,6} \and \\
   C.~Maitra\inst{7} \and 
   M.~Veresvarska\inst{8}
          }

   \institute{Leibniz-Institut für Astrophysik Potsdam (AIP), An der Sternwarte 16, 14482 Potsdam, Germany
        \and
             Institut für Astronomie und Astrophysik Tübingen, Universität Tübingen, Sand 1, Tübingen, 72076, Baden-Württemberg, Germany
        \and
             Potsdam University, Institute for Physics and Astronomy, Karl-Liebknecht-Straße 24/25, 14476 Potsdam, Germany
        \and
             South African Astronomical Observatory, PO Box 9, Observatory Road, Observatory 7935, Cape Town, South Africa
        \and
             Department of Astronomy, University of Cape Town, Private Bag X3, Rondebosch 7701, South Africa
        \and
             Department of Physics, University of the Free State, PO Box 339, Bloemfontein 9300, South Africa
        \and
             Max-Planck-Institut für Extraterrestrische Physik, Gießenbachstraße 1, 85748 Garching, Germany
        \and
             Department of Physics, Centre for Extragalactic Astronomy, Durham University, South Road, Durham DH1 3LE, UK
             }

   \date{Received 27 August 2025 / Accepted 4 November 2025}

 
  \abstract 
   {SRGt 062340.2-265751, a cataclysmic variable identified by \textit{SRG}/\ero\ thanks to its significant X-ray variability, remains poorly characterised despite the multi-wavelength follow-up. We present spectral and timing analyses from the first dedicated X-ray and ultraviolet observations with \xmm, complemented by \textit{SRG}/\ero\ data from four all-sky surveys (eRASS1-4) and ASAS-SN optical photometry. Our timing analysis reveals a >8$\sigma$ significant modulation at $3.6 \pm 0.5$ hours, likely representing the orbital period. Long-term ASAS-SN monitoring confirms the source as a VY Sculptoris-type nova-like system, while short-timescale X-ray and ultraviolet variability, down to a few minutes, suggests a possible underlying magnetic white dwarf. Two additional significant X-ray modulations at $43 \pm 1$ min and $36.0 \pm 0.7$ min tentatively point to the spin period of an intermediate polar. The best-fit \xmm\ energy spectra reveal a multi-temperature thermal plasma ($kT = 0.23, 0.94$, and $5.2$ keV), while the \textit{SRG}/\ero\ spectra are consistent with a single-temperature thermal plasma of a few keV. We estimate unabsorbed X-ray luminosities of \mbox{$\gtrsim 10^{32}$ erg s$^{-1}$} (0.2$-$12 keV). Broadband spectral energy distribution modelling, from near-ultraviolet to infrared, indicates a disc-dominated system consistent with a nova-like classification. We discuss these results in the context of the source's confirmed nova-like classification and its possible magnetic nature, a scenario increasingly supported by discoveries of intermediate polars exhibiting VY Sculptoris-type nova-like features.
    }


   \maketitle
%
\begin{table*}
\caption{X-ray observing log.}             
\label{log}      
\centering        
\resizebox{\textwidth}{!}{
\begin{tabular}{lccccccccccc}
\hline            
\rule{0pt}{2.5ex}Observatory & SKYTILE / ID & Epoch & Date & MJD & Exposure & Instrument & Mode & Filter & Net exposure & Rate \\
\rule{0pt}{1.5ex} &  &  & (yyyy-mm-dd) & & [s] &  &  &  & [s] & [cts s$^{-1}$] \\
\\[-2ex]
\hline
   \rule{0pt}{2.3ex}\textit{SRG} & 095108 & eRASS1 & 2020-04-07 & 58946.9 & 247 & \ero & Survey & FILTER & 107 & 0.74 $\pm$ 0.06 \\
   \rule{0pt}{2ex} & 095108 & eRASS2 & 2020-10-11 & 59133.4 & 236 & \ero & Survey & FILTER & 102 & 2.8 $\pm$ 0.1 \\
   \rule{0pt}{2ex} & 095108 & eRASS3 & 2021-04-05 & 59310.1 & 323 & \ero & Survey & FILTER & 141 & 2.53 $\pm$ 0.09 \\
   \rule{0pt}{2ex} & 095108 & eRASS4 & 2021-10-07 & 59494.6 & 356 & \ero & Survey & FILTER & 154 & 1.40 $\pm$ 0.06 \\
   \hline
   \rule{0pt}{2.3ex}\xmm & 0901220101 &  & 2023-04-21 & 60055.5 & 46000 & EPIC-pn & Small Window & THIN1 & 44139 & 1.075 $\pm$ 0.006 \\  
   \rule{0pt}{2ex} &  &  &  &  &  & EPIC-MOS1 & Small Window & THIN1 & 44626 &  0.236 $\pm$ 0.002  \\
   \rule{0pt}{2ex} &  &  &  &  &  & EPIC-MOS2 & Small Window & THIN1 & 44606 &  0.284 $\pm$ 0.003  \\
   \rule{0pt}{2ex} &  &  &  &  &  & RGS1 & Spec. HER$+$SES &  & 44830 &  0.049 $\pm$ 0.001  \\
   \rule{0pt}{2ex} &  &  &  &  &  & RGS2 & Spec. HER$+$SES &  & 44910 &  0.051 $\pm$ 0.001  \\
   \rule{0pt}{2ex} &  &  &  &  &  & OM & Fast & UVM2 & 41398 &  81.13 $\pm$ 0.04  \\
\hline
\end{tabular}} 
\end{table*}

\section{Introduction}

Cataclysmic variables (CVs), the most common type of accreting white dwarfs, are close binaries in which a white dwarf accretes matter from a low-mass companion star that fills its Roche lobe \citep[see][for a review]{Warner2003}. Depending on the strength of the white dwarf's magnetic field, these systems are typically classified as either magnetic (B $>$ 1 MG) or non-magnetic. Magnetic systems comprise polars and intermediate polars (IPs). In polars, the magnetic field is so strong (B $\ge$ 10 MG) that it prevents the formation of an accretion disc. Instead, matter is channelled along magnetic field lines and impacts the white dwarf near its magnetic poles, forming a shock-heated region that emits X-rays \citep[for a review, see][]{Cropper1990}. In these systems, the white dwarf's spin period ($P_{\rm spin}$) is typically synchronised with the orbital period \citep[$P_{\rm orb}$; see, for example,][]{Schwope2025}. IPs, by contrast, have weaker magnetic fields, allowing the formation of a truncated accretion disc. Therefore, matter is first accreted through the disc before being redirected along magnetic field lines to the poles, where it shocks and heats up similarly to polars. The post-shock plasma then cools via bremsstrahlung radiation, producing hard X-rays \citep[see the review by][]{Patterson1994}. Since the magnetic and rotational axes are usually misaligned, the X-ray emission is modulated at $P_{\rm spin}$. In these systems, $P_{\rm spin}$ and $P_{\rm orb}$ are not synchronised, and both modulations are typically identified through timing analysis, serving as a key diagnostic of their IP nature \citep[e.g.][]{Mukai2017}. In contrast, non-magnetic systems, which are mainly divided into dwarf novae and nova-likes, accrete matter through an accretion disc. Dwarf novae, the class from which the term cataclysmic variables originates, are transient systems that undergo outbursts when the disc becomes thermally unstable after accumulating enough matter \citep[see reviews by][]{Osaki1996,Hameury2020}. Nova-likes, on the other hand, accrete matter persistently at high rates \citep[$\dot{M}\sim10^{-8}M_{\odot}~yr^{-1}$,][]{Puebla2007,Ballouz2009} through a hot stable disc. In particular, a subclass of nova-likes known as VY Sculptoris (VY Scl) exhibit long-term variability in their optical light curves, including sporadic low states characterised by brightness drops of $\gtrsim$1 mag \citep[e.g.][]{King1998}. 

The \textit{Spektrum-Roentgen-Gamma} (\textit{SRG}) mission \citep[][]{Sunyaev2021} conducted 4.4 all-sky surveys between December 2019 and February 2022 using its two onboard instruments: the Mikhail Pavlinsky ART-XC \citep{Pavlinsky2021} and the extended ROentgen Survey with an Imaging Telescope Array \citep[\ero,][]{Predehl2021}. Following the release of the first \ero\ all-sky survey (eRASS1), the number of known X-ray sources reported in the literature increased by more than 60\% \citep{Merloni2024}. In particular, \ero\ has facilitated the discovery of new CV candidates that were subsequently confirmed through follow-up multi-wavelength observations, especially optical spectroscopy \citep[][]{Schwope2022b,Ok2023}.

SRGt 062340.2-265751 (hereafter \s) was identified during the second \ero\ all-sky survey (eRASS2), which revealed a factor-of-two increase in X-ray flux compared to eRASS1 \citep{Schwope2020b}. Its optical counterpart, with $G=12.4$ mag and a distance of $496\pm4$ pc, was identified in Gaia DR3 \citep[Gaia DR3 2899766827964264192;][]{Schwope2022}. Notably, this counterpart had already been reported in 2017 by Denis Denisenko and designated DDE 79 in the International Variable Star Index (VSX) catalogue\footnote{\url{https://vsx.aavso.org/index.php?view=detail.top&oid=477558}} of the American Association of Variable Star Observers (AAVSO) and by the Zwicky Transient Facility (ZTF19aaabzuh). Optical spectroscopy of the Gaia counterpart, obtained using the Wide Field Spectrograph (WiFeS) on the 2.3 m telescope at the Australian National University (ANU) and the High Resolution Spectrograph (HRS) on the 10 m Southern African Large Telescope (SALT), unequivocally identified the source as a CV. The SALT/HRS spectrum revealed the presence of the Bowen blend at $\sim$$\lambda4640$, a feature commonly observed in magnetic CVs \citep[e.g.][]{Schachter1991}. Archival serendipitous X-ray observations of the \s\ field, including \textit{ROSAT} (1RXS J062339.8-265744), \textit{XMM-Newton}, and \textit{Swift}/XRT (2SXPS J062339.9-265751), revealed a high degree of X-ray variability \citep{Schwope2022}, consistent with the \ero\ findings and suggestive of an IP \citep[e.g.][]{Norton1996}. However, this system's most puzzling property is its optical photometric variability. The Transiting Exoplanet Survey Satellite \citep[TESS;][]{Ricker2015} observed the source from 15 December 2018 to 6 January 2019, and again from 18 December 2020 to 13 January 2021. During the first epoch, a clear modulation at 3.9 h was detected \citep{Pichardo-Marcano2020}, yet this periodicity was absent in the second epoch \citep{Schwope2022}, which instead revealed a $\sim$25 min modulation. A period search in archival photometry from the All-Sky Automated Survey for Supernovae \citep[ASAS-SN;][]{Shappee2014,Hart2023}, confirmed the 3.9 h periodicity detected during the first TESS observation. Additionally, high-speed photometry obtained with the Sutherland High Speed Optical Camera \citep[SHOC;][]{Coppejans2013} on the 1 m telescope at the South African Astronomical Observatory (SAAO) revealed modulations at $\sim$35 min. The source is clearly variable, with potential periodicities at different wavelengths. However, despite the extensive datasets analysed, no periodic signal could be consistently confirmed across all observations. Given its unusual properties, \citet{Schwope2022} suggested that \s\ could be an X-ray underluminous IP, an overluminous non-magnetic CV, or perhaps a new type of CV.

To investigate the nature of the source and search for potential periodic modulations, we obtained a dedicated \textit{XMM-Newton} observation of \s, lasting 46 ks, in April 2023. In the following sections, we present an analysis of these new follow-up observations, along with the complete \ero\ dataset and archival broadband photometry from the near-UV to the IR. A companion paper by Brink et al. (in prep.) will present a detailed analysis of extensive optical photometry and spectroscopy.

\begin{figure*}
 \centering
 \includegraphics[trim=0mm 0mm 0mm 0mm,width=\textwidth]{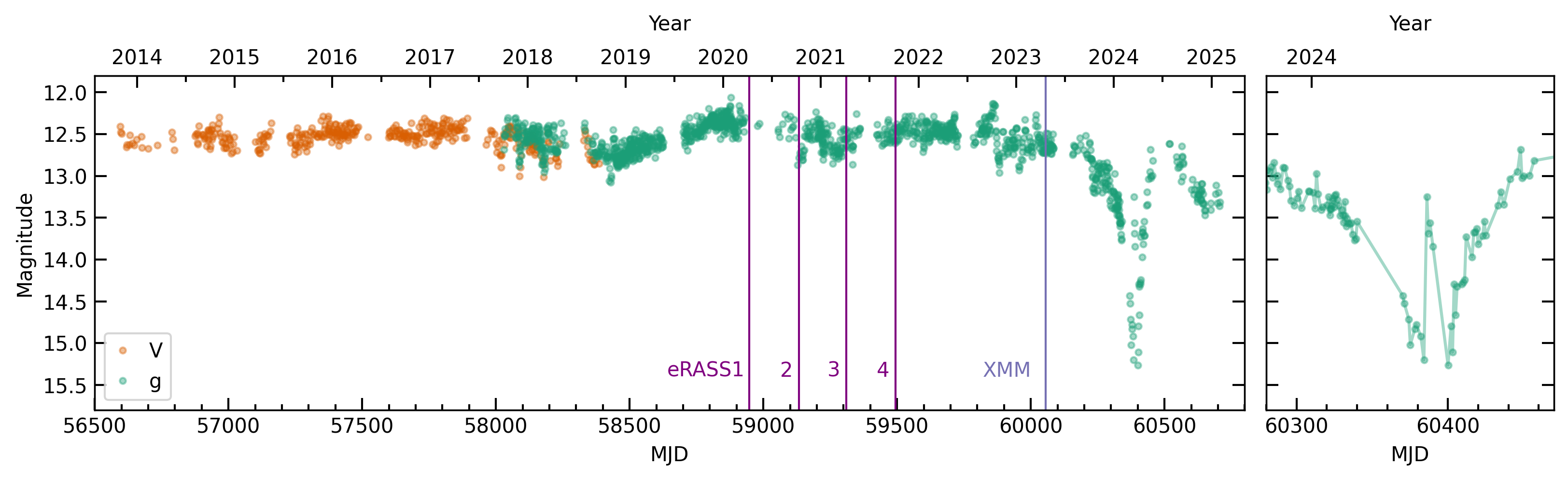}
 \caption{\textit{Left panel:} Long-term ASAS-SN light curve of \s, combining V- and g-band magnitudes. The purple vertical lines indicate the epochs of the \ero\ observations, while the lilac line marks the epoch of the \textit{XMM-Newton} observation. \textit{Right panel:} Detailed view of the light curve between MJD 60280 and MJD 60470.}
 \label{opt_lc}
\end{figure*}

\section{Observations}

\subsection{XMM-Newton}

We observed \s\ with the \textit{XMM-Newton Observatory} \citep{Jansen2001} on 21 April 2023 for a total exposure of 46 ks, using all onboard instruments (see Table \ref{log} for the observing log). Calibrated events lists and scientific data products were extracted using the Science Analysis System (\textsc{sas, v21.0.0}) in combination with \textsc{heasoft v6.33.1} and the latest calibration files, following standard data reduction procedures.\footnote{\url{https://www.cosmos.esa.int/web/xmm-newton/sas-threads}}

We processed the data collected by the pn detector \citep{Struder2001} and the Metal Oxide Semi-conductor \citep[MOS1 and MOS2;][]{Turner2001} detectors of the European Photon Imaging Camera (EPIC), using the \textsc{sas} tasks \texttt{epproc} and \texttt{emproc}, respectively. Source events were extracted using a circular region with a radius of 30$\arcsec$ in all detectors. Background events were selected from two rectangular regions within the EPIC-pn window, carefully avoiding the readout direction, and from two circular regions with a radius of 50$\arcsec$ in CCDs adjacent to the source CCD in the EPIC-MOS data. We filtered the event files to exclude periods of background flaring using a sigma-clipping algorithm with a 3$\sigma$ threshold, and used \texttt{tabgtigen} to define the corresponding good time intervals. We note, however, that the observation was barely affected by flaring, with only 500 s and 1000 s removed from the EPIC-pn and EPIC-MOS2 datasets, respectively, while no filtering was required for EPIC-MOS1. Additional cleaning was applied using the recommended filters: \texttt{FLAG = 0} and \texttt{PATTERN $\leq$ 4} for EPIC-pn, and \texttt{FLAG = 0} and \texttt{PATTERN $\leq$ 12} for EPIC-MOS. A barycentric correction was then applied to the final event lists using the \texttt{barycen} task with DE200 ephemeris. Background-corrected light curves were produced using \texttt{evselect} and \texttt{epiclccorr}. Following standard procedures, we also generated source and background spectra, as well as response matrix files (RMFs) and ancillary response files (ARFs). The spectra were rebinned using \texttt{specgroup} to ensure a minimum of 25 counts per spectral bin, to avoid oversampling the full width at half-maximum of the energy resolution by more than a factor of three, and to link all the spectral files consistently.

We used the \textsc{sas} task \texttt{rgsproc} to process the data from the Reflection Grating Spectrometers \citep[RGS1 and RGS2;][]{denHerder2001} and obtain the data products. The sigma-clipping algorithm was applied again to check for flaring events on CCD 9, which is the closest to the telescope's optical axis and the most sensitive to background flares. However, no enhanced activity was detected. When analysing RGS data, it is important to note that one CCD chip in each detector failed early in the mission due to electronic issues, affecting the spectral ranges 11$-$14 $\mathring{\mathrm{A}}$ ($\sim$0.9$-$1.1 keV) in RGS1 and 20$-$24 $\mathring{\mathrm{A}}$ ($\sim$0.5$-$0.6 keV) in RGS2. Finally, we used \texttt{rgscombine} to merge the RGS1 and RGS2 spectra and \texttt{specgroup} to rebin the spectra to a minimum of 25 counts per bin, set an oversampling factor of 3, and link the spectral files.

The Optical Monitor \citep[OM;][]{Mason2001} was operated in fast mode using the UVM2 filter ($\lambda_{\rm eff} = 2310~\mathring{\mathrm{A}}$). We generated a light curve using the \texttt{omfchain} task, and then applied the barycentric correction using the \texttt{barycen} task with DE200 ephemeris.

\subsection{SRG/eROSITA}

The field of \s\ was observed during four all-sky scans (eRASS1-4) using the \ero\ instrument aboard the \textit{SRG} mission, as part of the \ero\ All-sky Survey \citep{Merloni2024}. These observations, with a total exposure of \mbox{$\sim$1.1 ks}, span approximately 1.5~years (from April 2020 to October 2021; see the observing log in Table \ref{log}) and allow us to probe the long-term evolution of \s\ in the \mbox{0.2$-$8 keV} X-ray band. In accordance with the eROSITA scanning pattern, each all-sky scan consists of seven to nine slews over the field, separated by four hours (one eroday), which is the time required for the instrument to complete one full rotation \citep{Merloni2024}.  

We obtained event lists containing detected X-ray photons from all seven X-ray telescopes in \ero, which were processed with the latest pipeline (version c030). These were cleaned using the eROSITA Science Analysis Software System \citep[eSASS; user version 240410.0.4,][]{Brunner2022}, retaining only photons with valid patterns (\texttt{PATTERN $\leq$ 15}) and standard good quality flagging criteria \citep[\texttt{FLAG} = 0xE000F000; e.g. see section 3.2.2 in][for more details]{Kurpas2024}. We also removed high-background flaring periods by applying the \texttt{flaregti} task to create updated good time interval (GTI) lists, which were then applied to the event lists. To identify \s\ and nearby field sources, we conducted a source detection within 20\arcmin\ of \s's position in the \mbox{0.2$-$2.3 keV} band, following the methods of \citet{Brunner2022} and \citet{Merloni2024}. Based on these results, we applied the \texttt{srctool} task to define optimised source and background regions, extract source spectra, and generate light curves. The spectra were binned to contain a minimum of 1 and 25 counts per spectral bin using the \texttt{grppha} task. The light curve was binned into 1 eroday intervals (see above for definition), ensuring one data point per slew containing photons from \s.

\begin{figure*}
 \centering
 \includegraphics[trim=0mm 0mm 0mm 0mm,width=0.94\textwidth]{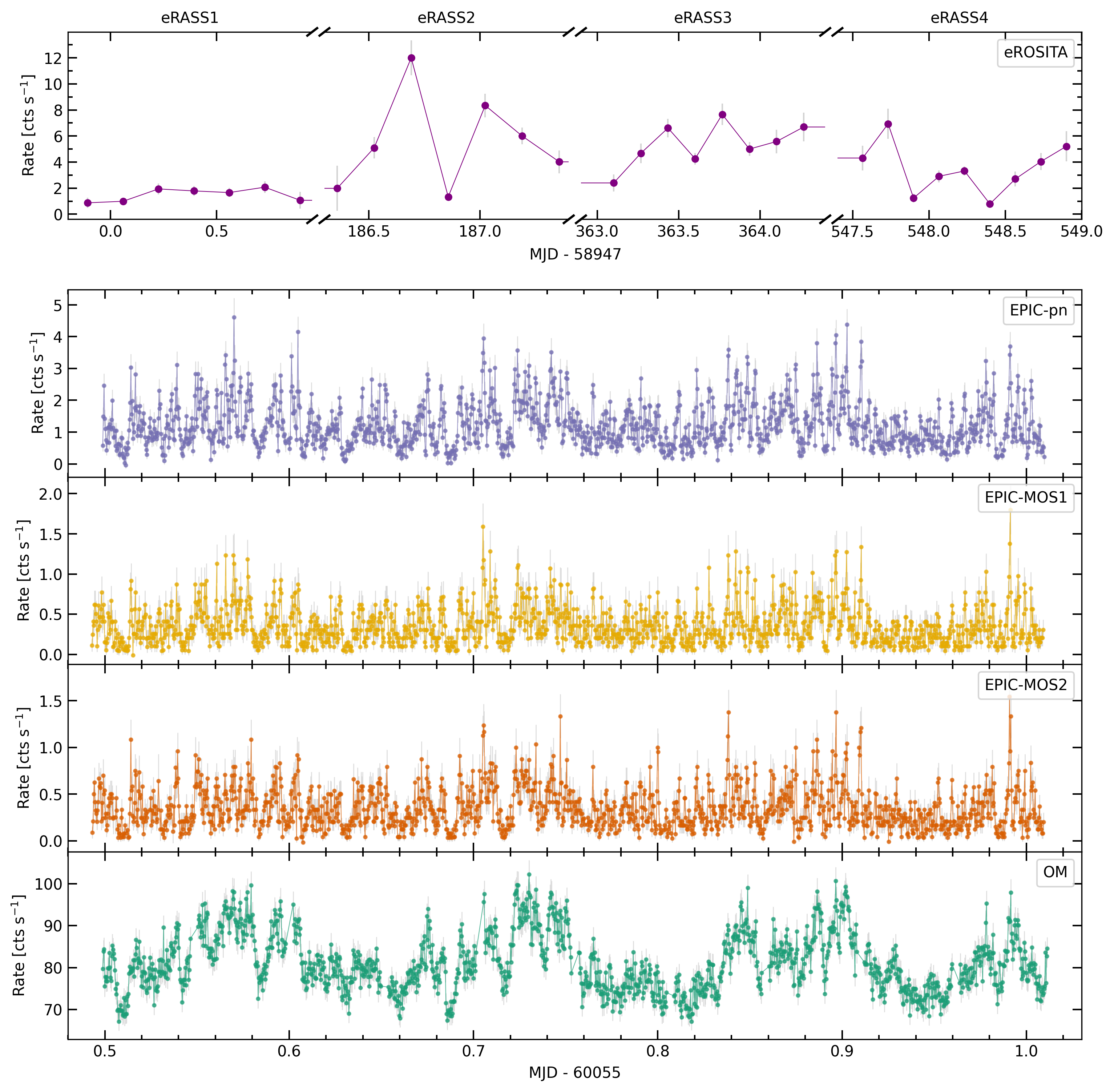}
 \caption{\textit{Top panel:} \ero\ light curve of \s\ in the 0.2$-$8 keV energy band across the four all-sky surveys (eRASS1-4). Each data point represents the mean rate of a single scan over the source position, separated by an eroday (4 h; see text for more details). \textit{Second to fifth panels:} \textit{XMM-Newton}/EPIC-pn, EPIC-MOS1, and EPIC-MOS2 30 s binned light curves of \s\ in the 0.2$-$12 keV energy band, followed by the OM 30 s binned light curve obtained with the UVM2 filter.}
 \label{lcs}
\end{figure*}

\section{Analysis and results}

\subsection{Long-term optical variability}

To analyse the optical evolution of \s, we extracted a long-term ASAS-SN light curve using the ASAS-SN Sky Patrol V2.0.\footnote{\url{http://asas-sn.ifa.hawaii.edu/skypatrol/}} The resulting optical light curve, shown in the left panel of Fig. \ref{opt_lc}, exhibits significant variability. Between November 2013 (MJD $\sim$ 56600) and June 2023 (MJD $\sim$ 60100) the V/g-band magnitudes oscillated between $\sim$13 and 12 mag. Our \ero\ and \textit{XMM-Newton} observations were conducted during this period and are indicated in Fig. \ref{opt_lc}. The most dramatic variability, however, occurred from September 2023 (MJD $\sim$ 60200), when the source exhibited two drops in brightness. The first was a sharp $\sim$3~mag decline to $g\simeq15.2$, interrupted by a brief rebrightening around MJD $\sim60390$ (duration $<20$~days) that reached $g\simeq13.5$ (see the right panel of Fig. \ref{opt_lc}), after which the source settled again near $g\simeq15.3$. It then recovered to approximately its pre-dip level ($g\simeq12.6$) before a second $\sim$1~mag decline to $g\simeq13.5$. Such low states are characteristic of the VY Scl subclass of nova-likes \citep{King1998,Warner2003}, firmly establishing \s\ as a member of this group. The rebrightening during the first drop resembles the dwarf-nova-type outbursts reported during the low states of a few VY Scl systems \citep[e.g.][]{Pavlenko1999}; however, such rebrightenings are rare \citep[e.g.][]{Hameury2020}.

\subsection{X-ray and ultraviolet timing analysis}
\label{timing}

The top panel of Fig. \ref{lcs} shows the \ero\ light curve of \s\ in the 0.2$-$8 keV energy band across the four all-sky surveys (eRASS1-4). In addition to the pronounced variability between eRASS1 and eRASS2, which led to the detection of the source, significant variability is evident throughout the dataset, both on long timescales (from one survey to the next, over $\sim$6 months) and on short timescales (from one scan of the source position to the next, over $\sim$4 h). To quantify the overall variability, we computed the maximum observed rate deviation, following \citet[][see also \citealt{Traulsen2020}]{Kurpas2024}:

\begin{equation}
\text{RATE\_VAR} = \max_{k,l \in [1,n]} 
\frac{|R_k - R_l|}{\sqrt{\sigma_k^2 + \sigma_l^2}}.
\label{eq:rate_var}
\end{equation}

\noindent Here $R_k$ and $R_l$ are the count rates at epochs $k$ and $l$, and $\sigma_k$ and $\sigma_l$ are their corresponding uncertainties. The most significant variability (exceeding $8\sigma$ and corresponding to a rate change of more than 90\%) occurs between MJD 59133.69 (eRASS2) and MJD 59495.40 (eRASS4).

\begin{table}
\caption{Summary of the peak values of the >5$\sigma$ significant modulations detected in the Lomb-Scargle power spectra shown in Fig. \ref{ls}.}             
\label{peaks}      
\centering        
\resizebox{\columnwidth}{!}{
\begin{tabular}{l c c c c c}
\hline            
\rule{0pt}{2.5ex} & LS frequency & Gauss fit & Gauss fit & EPIC-pn & OM \\
\rule{0pt}{1.5ex} & [$\times10^{-4}$ Hz] & [$\times10^{-4}$ Hz] & [Time] & sig. & sig. \\
\\[-2ex]
\hline     
   \rule{0pt}{2.3ex}EPIC-pn & 0.760 & 0.8 $\pm$ 0.1 & 3.6 $\pm$ 0.5 h & 8$\sigma$ & >8$\sigma$ \\
   \rule{0pt}{2ex} & 3.911 & 3.9 $\pm$ 0.1 & 43 $\pm$ 1 min & 5$\sigma$ & 1$\sigma$ \\
   \rule{0pt}{2ex} & 4.636 & 4.63 $\pm$ 0.09 & 36.0 $\pm$ 0.7 min & 8$\sigma$ & 3$\sigma$ \\
   \rule{0pt}{2ex} & 7.856 & 7.9 $\pm$ 0.1 & 21.2 $\pm$ 0.3 min & 6$\sigma$ & 3$\sigma$ \\
   \rule{0pt}{2ex} & 9.761 & 9.76 $\pm$ 0.09 & 17.1 $\pm$ 0.2 min & 5$\sigma$ & -- \\
\hline
   \rule{0pt}{2.3ex}OM & 0.759 & 0.8 $\pm$ 0.1 & 3.6 $\pm$ 0.5 h & 8$\sigma$ & >8$\sigma$ \\
   \rule{0pt}{2ex} & 1.371 & 1.37 $\pm$ 0.08 & 2.0 $\pm$ 0.1 h & <1$\sigma$ & $8\sigma$ \\
   \rule{0pt}{2ex} & 1.905 & 1.9 $\pm$ 0.1 & 1.4 $\pm$ 0.1 h & 2$\sigma$ & $5\sigma$ \\
   \rule{0pt}{2ex} & 6.030 & 6.02 $\pm$ 0.09 & 27.7 $\pm$ 0.4 min & <1$\sigma$ & 5$\sigma$ \\
   \hline   
\end{tabular}} 
\end{table}

In the second to fourth panels of Fig. \ref{lcs}, we present 30 s binned light curves extracted from the EPIC-pn, EPIC-MOS1, and EPIC-MOS2 photon event lists in the 0.2$-$12 keV energy range. The bin size was chosen to avoid empty bins while maintaining high cadence. In the bottom panel of Fig. \ref{lcs}, we also show the ultraviolet (UV) OM \mbox{30 s} binned light curve. Clear variability is observed in all the \xmm\ light curves, with notable similarities between the different datasets.

\begin{figure}
    \centering
    \includegraphics[width=\columnwidth]{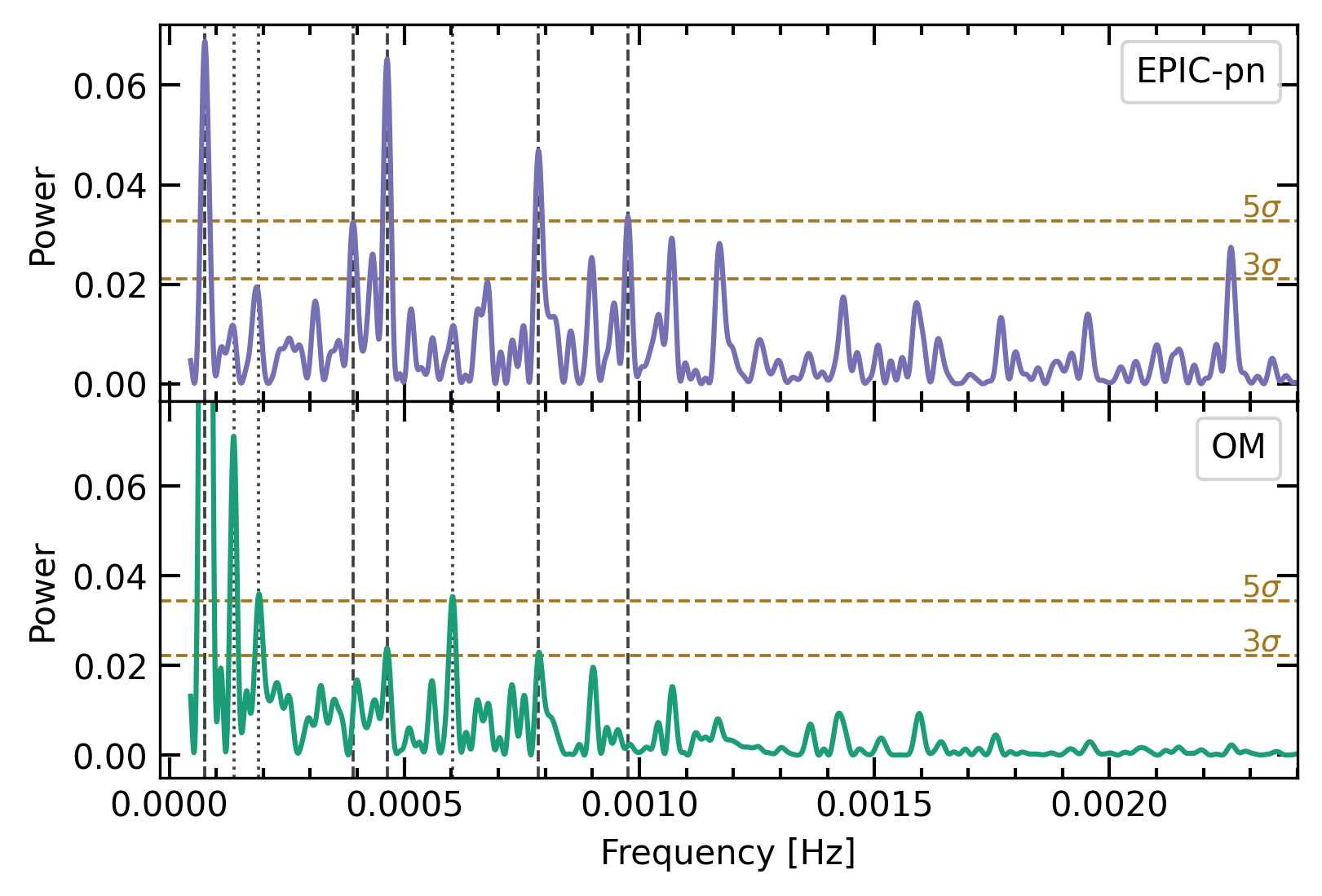}
    \caption{Lomb-Scargle power spectra of the EPIC-pn light curve in the 0.2$-$12 keV energy band (top) and the OM light curve obtained with the UVM2 filter (bottom). The dashed horizontal lines indicate the 3$\sigma$ and 5$\sigma$ power significance levels. The grey vertical lines (dashed for EPIC-pn and dotted for OM) denote the frequency of the peaks with a significance higher than 5$\sigma$. The frequencies and significance levels of the marked peaks are summarised in Table \ref{peaks}.}
    \label{ls}
\end{figure}

Since EPIC-pn and OM produced the highest count rates, we focused our timing analysis on these instruments to search for periodic modulations. We used the Lomb-Scargle algorithm \citep{Lomb1976,Scargle1982} to search for periodic modulations within the range $60-22000$ s, defined by twice the bin size (corresponding to the Nyquist frequency) and approximately half of the duration of the observation (corresponding to the longest detectable modulation). Fig. \ref{ls} shows the resulting power spectra, revealing significant power peaks at several frequencies. We marked the peak values of the >5$\sigma$ significant modulations in Fig. \ref{ls} and listed them in Table \ref{peaks}. All significance levels indicate the probability that the observed power is not due to white noise, with the likelihood of peaks arising from white noise estimated through the false-alarm probability \citep{Baluev2008}. The errors correspond to half the full width at half-maximum of a Gaussian fit to each peak. Table \ref{peaks} also includes the significance levels of the modulations in the other power spectrum (EPIC-pn for OM, and vice versa). The only modulation detected with high significance by both EPIC-pn and OM is the highest peak in the power spectra at $\sim7.6\times10^{-5}$ Hz ($3.6\pm0.5$ h), which we tentatively interpret as the orbital period (see Section \ref{modulations} for further discussion). We analysed possible correlations between the significant modulations to establish their origin. We found that if P$_{orb} = 3.6$~h, the modulations at 2 and 1.4~h are likely the first and second harmonics, respectively, within uncertainties on all values. Additionally, we identified two potential candidates for a spin period that provide meaningful relations to the other significant modulations: 36 min, the second strongest peak in the EPIC-pn power spectrum and nearly as significant as the 3.6 h peak, and 43 min. We note, however, that these interpretations should be treated with caution until further evidence clarifies their nature. Table \ref{corr} summarises the correlations between the modulations listed in Table \ref{peaks} for these two hypothetical cases. Fig. \ref{foldedLCs} presents the EPIC-pn and OM light curves folded on the 3.6~h, 43~min, and 36~min modulations. The 43~min signal shows no clear modulation in either dataset, yet its pulse fraction is comparable to that of the more evident 36~min modulation: $24\%$ for EPIC-pn and $2.4-2.8\%$ for OM. In both cases, the OM folded light curve exhibits very low pulse fractions, reflecting the weak presence of these signals at ultraviolet wavelengths.

\begin{table}
\caption{Summary of the correlations between the modulations listed in Table \ref{peaks}, assuming P$_{orb} = \Omega^{-1} = 3.6$ h and two possible P$_{spin} = \omega^{-1}$.}
\label{corr}      
\centering        
\begin{tabular}{l c c}
\hline
\rule{0pt}{2.5ex} & Frequency & Period \\
\\[-2ex]
\hline
   \rule{0pt}{2.3ex}If P$_{spin} = 36$ min & $\omega - \Omega$ & 43 min \\
   \rule{0pt}{2ex} & $2(\omega - \Omega)$ & 21.2 min \\
\hline
   \rule{0pt}{2.3ex}If P$_{spin} = 43$ min & $\omega-\Omega$ & 53.7 min\tablefootmark{(*)} \\
   \rule{0pt}{2ex} & $\omega+\Omega$ & 36 min \\
   \rule{0pt}{2ex} & $2(\omega-\Omega)$ & 27.7 min \\
   \rule{0pt}{2ex} & $2\omega$ & 21.2 min \\
   \hline   
\end{tabular}
\tablefoot{\tablefoottext{*}{Although not listed in Table \ref{peaks}, this period corresponds to a peak in the EPIC-pn power spectrum with a significance below 3$\sigma$.}}
\end{table}

\begin{figure*}
    \centering
    \includegraphics[width=\textwidth]{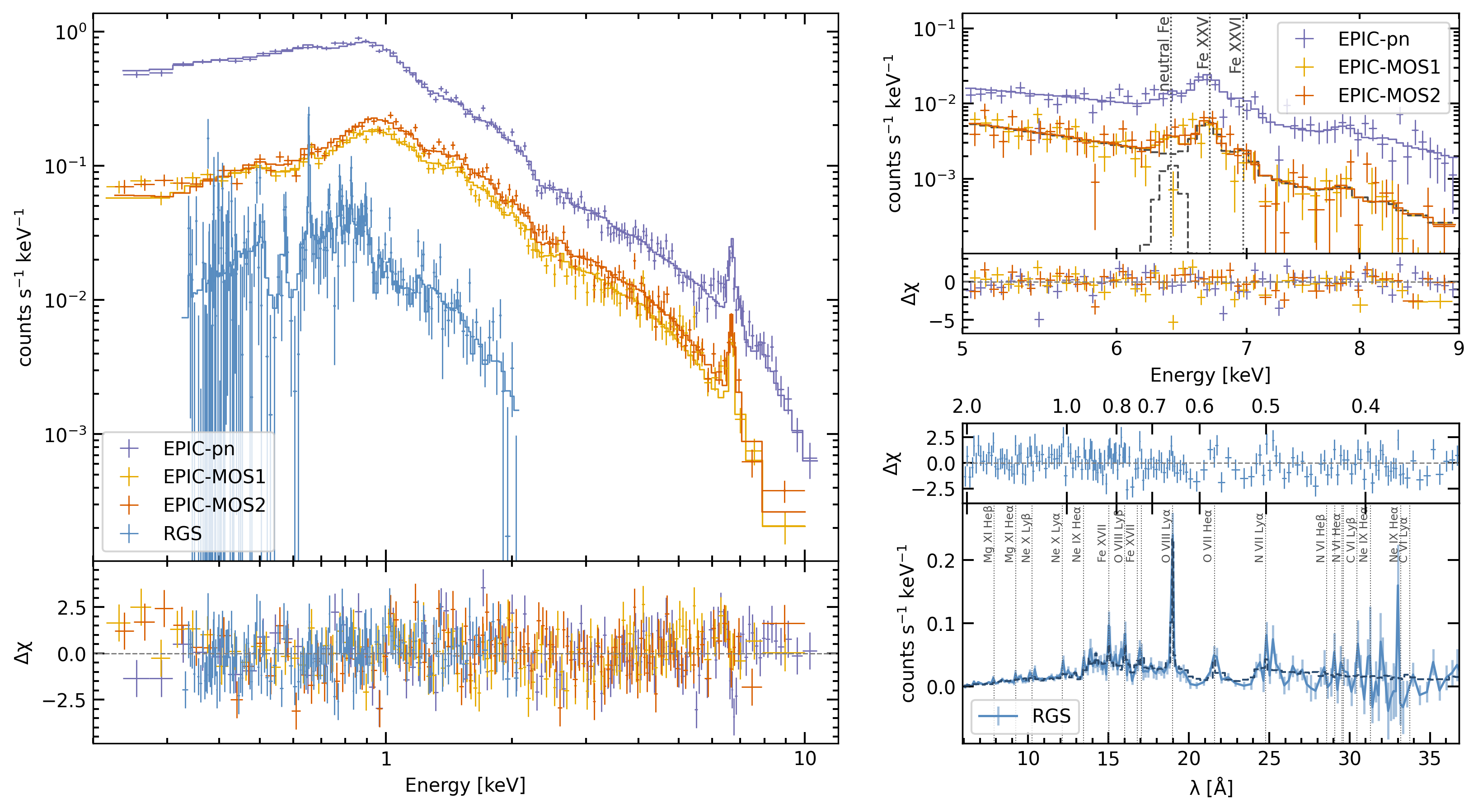}
    \caption{\textit{Left panel:} \xmm\ energy spectra of \s, shown with the best-fitting model (\texttt{TBabs*(apec+apec+apec)}). The bottom panel shows the normalised residuals. The best-fit parameters are listed in Table \ref{spec_param}. \textit{Top right panel:} Zoomed-in image of the Fe K complex, fitted with a \texttt{TBabs*(apec+gaussian)} model. The dashed lines represent the model components for the EPIC-MOS1 spectrum, shown as an example. The vertical dotted lines mark the expected energies of the different Fe-line components. \textit{Bottom right panel:} RGS spectrum and best-fitting model, identical to those shown in the left panel, but plotted as a function of wavelength. The vertical lines indicate several prominent and commonly observed transitions.}
    \label{xmm_spec}
\end{figure*}

We also computed the power spectra of the EPIC-pn data across multiple energy bands (0.2$-$0.5, 0.5$-$1, 0.5$-$3, 1$-$3, and 3$-$12 keV). However, these were less significant than those derived from the full energy band and did not provide any additional information. Nevertheless, the modulation we associate with $P_{\rm orb}$ was the only one consistently significant ($>3\sigma$) across all analysed bands. We further examined the power spectra of the full-band (0.2$-$12 keV) EPIC-pn and OM observations by splitting each into two halves. Although the 3.6 h modulation could no longer be detected due to the reduced time interval, the shorter periodicities detected in the full observations were present in the first half but fell below 3$\sigma$ significance in the second half for both EPIC-pn and OM. This transient behaviour is consistent with the variability previously reported for this source \citep[e.g. in TESS datasets;][]{Schwope2022}.

In addition to the periodic modulations discussed above, the \xmm\ light curves (Fig. \ref{lcs}) show clear short-term variability. To quantify this, we again applied Eq. \ref{eq:rate_var}, measuring the maximum rate deviation over 5 min intervals. For the EPIC-pn light curve, the median significance across all 5 min intervals is $3.8\sigma$, with individual intervals reaching $\sim6\sigma$ and count rate changes of $\sim90\%$ on timescales of only a few minutes. To assess whether such flickering could be attributed solely to absorption, we simulated spectra with varying Galactic column densities ($N_{\rm H}$). Adopting an absorbed thermal plasma model (\texttt{tbabs*apec}) with \mbox{$kT = 8$ keV}, we found that $N_{\rm H}$ would need to increase by roughly three orders of magnitude to produce a 90\% drop in count rate. We therefore conclude that the flickering is primarily due to intrinsic variability in the source emission, rather than absorption effects. Applying the same methodology to the OM light curve over 5 min intervals yields a lower median variability significance of $2.4\sigma$.

\subsection{X-ray spectral analysis}
\label{spec_ana}

The left panel of Fig. \ref{xmm_spec} shows the \xmm\ spectra from the EPIC-pn, EPIC-MOS1, EPIC-MOS2, and RGS detectors. The EPIC spectra clearly exhibit the Fe K complex at 6$-$7~keV (see the top right panel of Fig. \ref{xmm_spec} for details), while the RGS spectrum reveals several emission lines at low energies (see the bottom right panel of Fig. \ref{xmm_spec} and the text below for details). We analysed all spectra using \textsc{xspec} \citep[v.12.14.0b;][]{Arnaud1996}, modelling interstellar absorption with the Tuebingen-Boulder model (\texttt{TBabs}) and adopting solar abundances from \citet{Wilms2000}. A simultaneous fit was performed for all four spectra, linking model parameters across instruments and using $\chi^2$ statistics. To account for calibration uncertainties, we included the \texttt{constant} scaling factor, left free to vary between instruments. A single \texttt{apec} model, representing a collisionally ionised plasma in thermal equilibrium, failed to adequately reproduce the data ($\chi^{2}_{\nu}(\nu)=4.6(536)$). Adding second and third \texttt{apec} components significantly improved the fit, suggesting emission from a cooling, multi-temperature plasma. Attempts to use multi-temperature isobaric cooling-flow models (\texttt{cevmkl} and \texttt{mkcflow}) or a partial covering absorber \citep[\texttt{pcfabs}; see][]{Norton1989}, often used in IPs to model X-ray absorption in the accretion column, did not improve the fit. We therefore adopted an absorbed three-component thermal plasma model (\texttt{TBabs*(apec+apec+apec)}) as the best fit. This model is shown in the left panel of Fig. \ref{xmm_spec}, and the best-fitting parameters with 1$\sigma$ uncertainties are listed in Table \ref{spec_param}. We derived a Galactic column density of \mbox{$N_{\rm H} = 3.4^{+0.3}_{-0.2} \times 10^{20}$ cm$^{-2}$}, slightly lower than the value of \mbox{$4.08 \times 10^{20}$ cm$^{-2}$} used by \citet{Schwope2022}. Using the \texttt{cflux} convolution model, we obtained an unabsorbed flux of \mbox{$3.13\times10^{-12}$ erg cm$^{-2}$ s$^{-1}$} in the \mbox{0.2$-$12 keV} band, corresponding to a luminosity of \mbox{$9.2\times10^{31}$ erg s$^{-1}$}, assuming a distance of \mbox{$496\pm4$ pc} \citep{Schwope2022}. 

\begin{table*}
\caption{Best-fitting parameters for the four \ero\ epochs using the \texttt{TBabs*apec} model, and from the simultaneous fit to all \xmm\ spectra using the \texttt{TBabs*(apec+apec+apec)} model.}   
\label{spec_param}      
\centering        
\begin{tabular}{llccccc}
\hline            
\rule{0pt}{2.5ex}Parameter &  & eRASS1 & eRASS2 & eRASS3 & eRASS4 & \xmm\ \\
\\[-2ex]
\hline     
   \rule{0pt}{2.3ex}$N_{\rm H}$ & [$10^{20}$ cm$^{-2}$] & 3.4 (fixed) & 3.4 (fixed) & 3.4 (fixed) & 3.4 (fixed) & $3.4^{+0.3}_{-0.2}$ \\
   \rule{0pt}{2.3ex}$kT_1$ & [keV] & $2.3^{+2.1}_{-0.5}$ & $13^{+29}_{-5}$ & $9^{+14}_{-3}$ & $4.8^{+1.3}_{-0.9}$ & $0.23 \pm 0.01$ \\
   \rule{0pt}{2.3ex}$kT_2$ & [keV] & -- & -- & -- & -- & $0.94^{+0.01}_{-0.02}$ \\
   \rule{0pt}{2.3ex}$kT_3$ & [keV] & -- & -- & -- & -- & $5.2 \pm 0.1$ \\
   \rule{0pt}{2.3ex}$\chi^{2}_{\nu}(\nu)$\tablefootmark{(a)} &  & 117.51(127) & 0.88(22) & 0.94(26) & 0.83(16) & 1.35(532) \\
   \rule{0pt}{2.3ex}Flux$_{\rm ~0.2-12~keV}$\tablefootmark{(b)} & [$10^{-12}$ erg cm$^{-2}$ s$^{-1}$] & $2.0 \pm 0.2$ & $17.4 \pm 0.7$ & $13.8 \pm 0.5$ & $6.1 \pm 0.3$ & $3.13 \pm 0.01$ \\
   \rule{0pt}{2.3ex}Luminosity$_{\rm ~0.2-12~keV}$\tablefootmark{(c)} & [$10^{32}$ erg s$^{-1}$] & $0.59 \pm 0.05$ & $5.1 \pm 0.2$ & $4.1 \pm 0.2$ & $1.79 \pm 0.09$ & $0.92 \pm 0.02$ \\
   \rule{0pt}{2.3ex}Luminosity$_{\rm ~0.25-10~keV}$\tablefootmark{(c)} & [$10^{32}$ erg s$^{-1}$] & $0.57 \pm 0.05$ & $4.7 \pm 0.2$ & $3.8 \pm 0.2$ & $1.71 \pm 0.09$ & $0.88 \pm 0.01$ \\   
   \rule{0pt}{2.3ex}Luminosity$_{\rm ~0.01-100~keV}$\tablefootmark{(c)} & [$10^{32}$ erg s$^{-1}$] & $0.63 \pm 0.06$ & $5.3 \pm 0.2$ & $4.2 \pm 0.2$ & $1.9 \pm 0.1$ & $1.05 \pm 0.02$ \\   
   \hline   
\end{tabular}
\tablefoot{Uncertainties are given at the 1$\sigma$ confidence levels.
\tablefoottext{a}{For eRASS1, the value corresponds to the C-statistic (d.o.f).}
\tablefoottext{b}{Unabsorbed fluxes.} 
\tablefoottext{c}{Unabsorbed luminosities were estimated assuming a distance of $496\pm4$ pc.}
}
\end{table*}

A closer inspection of the \mbox{Fe K$\alpha$} complex (top right panel of Fig. \ref{xmm_spec}) reveals, in addition to the expected \ion{Fe}{XXVI} (H-like) line at \mbox{6.97 keV} and \ion{Fe}{XXV} (He-like) line at \mbox{6.7 keV}, an emission excess corresponding to the neutral fluorescent \mbox{Fe K$\alpha$} line at \mbox{6.4 keV}. While the \ion{Fe}{XXVI} and \ion{Fe}{XXV} lines are typical of emission from hot, ionised plasma, the \mbox{6.4 keV} line likely indicates reflection off the surface of the white dwarf \citep[e.g.][]{Hellier2004,Rana2006}, and possibly also off the accretion disc \citep[][]{Done1997,Ishida2009}. The spectra shown in the top right panel of Fig. \ref{xmm_spec} were grouped to a minimum of 1 count per bin and analysed using the c-statistic. We initially fitted the \mbox{5$-$9 keV} energy range with an absorbed thermal plasma plus a Gaussian component fixed at \mbox{6.4 keV} to model the fluorescent line (\texttt{TBabs*(apec+gaussian)}). The Galactic column density was fixed to the value obtained from the simultaneous fit to \xmm\ spectra (\mbox{$N_{\rm H} = 3.4 \times 10^{20}$ cm$^{-2}$}). As the Gaussian line width was poorly constrained, we next fitted only the thermal plasma component (\texttt{TBabs*apec}) excluding the \mbox{6.2$-$6.45 keV} interval to account only for the ionised components. This gave a plasma temperature of \mbox{$kT = 6.5\pm0.5$ keV}. We then refitted the full \mbox{5$-$9 keV} range, again including the Gaussian, this time fixing the plasma temperature at \mbox{6.5 keV}. This fit yielded a Gaussian line width of \mbox{$\sigma = 0.15\pm0.06$ keV}, consistent with the EPIC energy resolution (of \mbox{150 eV} at \mbox{6.4 keV}). We therefore repeated the fit with the Gaussian width fixed at zero while allowing the plasma temperature to vary, obtaining \mbox{$kT = 6.6\pm0.5$ keV}, consistent with the previous value and indicative of a plasma hotter than the \mbox{5.2 keV} derived from the full \mbox{0.2$-$12 keV} fit. From this fit we derived an equivalent width (EW) for the \mbox{6.4 keV} line of \mbox{$150^{+60}_{-50}$ eV}.\footnote{While EWs are conventionally negative for emission lines, we report the absolute value for clarity.}

The bottom right panel of Fig. \ref{xmm_spec} shows the RGS spectrum in wavelength units for clarity, along with the best-fitting model (\texttt{TBabs*(apec+apec+apec)}) from the simultaneous fit to the \xmm\ spectra (see left panel of Fig. \ref{xmm_spec} and Table \ref{spec_param}). Several prominent emission lines are visible, primarily from oxygen and iron, and possibly also from nitrogen, neon, carbon, and magnesium. The strongest oxygen features (H-like \ion{O}{VIII} at \mbox{18.97 $\mathring{\mathrm{A}}$} \mbox{(0.65 keV)} and the He-like \ion{O}{VII} triplet at \mbox{$\sim$21.6, 21.8, 22.1 $\mathring{\mathrm{A}}$} \mbox{($\sim$0.574, 0.568, 0.561 keV)}) are well reproduced by our multi-temperature collisionally ionised plasma model. In contrast, several other features (e.g. \ion{C}{VI}, \ion{N}{VI}, and \ion{Ne}{IX}) are not accounted for by the model and may instead originate from photoionisation \citep{Mukai2003,Girish2007}, which can occur in low-density environments or under sufficiently strong radiation fields; for \s, an intense extreme-UV radiation field is the likely cause. However, even the well-modelled oxygen lines are insufficiently constrained (e.g. the individual components of the \ion{O}{VII} triplet are blended) to permit reliable diagnostics from line ratios, such as temperature or density estimates. 

\begin{figure*}
    \centering
    \includegraphics[width=\textwidth]{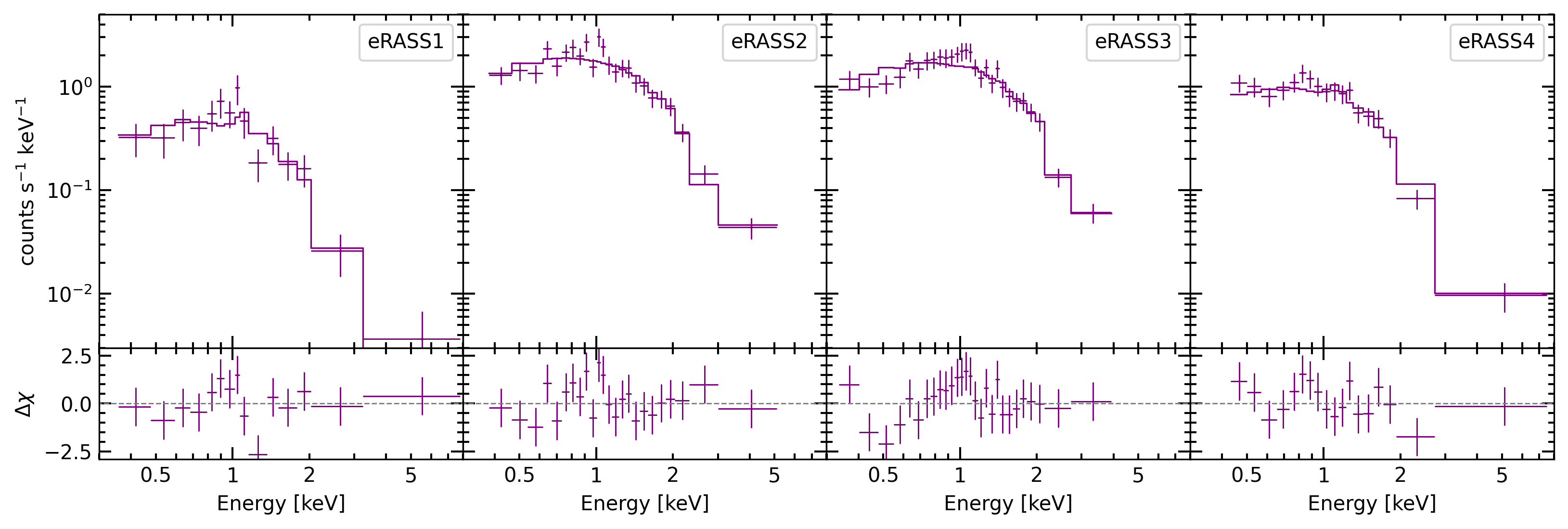}
    \caption{\ero\ energy spectra of \s\ from the four observing epochs, shown with the best-fitting model (\texttt{TBabs*apec}). For clarity, we show the eRASS1 spectrum rebinned to 10 counts per bin, although the analysis was performed using a 1 count per bin spectrum (see text). For eRASS2-4, we show the 25 counts per bin spectra used in the analysis. Normalised residuals are shown in the bottom panels. The best-fitting parameters are listed in Table \ref{spec_param}.}
    \label{eros_spec}
\end{figure*}

To analyse the energy spectra from the four \ero\ observing epochs, we used spectra grouped to 25 counts per bin for eRASS2-4 and applied $\chi^2$ statistics, while for eRASS1 we used a 1 count per bin spectrum (due to its lower number of counts) and applied the C-statistic. In all cases, the spectra were well described by a single absorbed thermal plasma model (\texttt{TBabs*apec}), with no significant residuals. We fixed the Galactic column density to the value obtained from the \xmm\ spectral analysis ($3.4\times10^{20}$ cm$^{-2}$) and derived 1$\sigma$ uncertainties for all fit parameters. The resulting spectra and best-fitting models are shown in Fig. \ref{eros_spec}, and the corresponding parameters are listed in Table \ref{spec_param}. The derived plasma temperatures span 2.3$-$13 keV (with large uncertainties) and are broadly comparable to the hottest \xmm\ component \mbox{($5.2\pm0.1$ keV)}. As an additional test, we fitted all four \ero\ spectra simultaneously with a single absorbed plasma model, obtaining \mbox{$kT = 6.5^{+1.1}_{-0.8}$ keV}, slightly higher than the hottest \xmm\ temperature but of the same order. The unabsorbed \mbox{0.2$-$12 keV} fluxes derived for each epoch imply luminosities in the range $0.59-5.1\times10^{32}$ erg s$^{-1}$ (assuming a distance of 496 pc; see Table \ref{spec_param}), consistent with the long-term variability observed in the \ero\ light curve (top panel of Fig. \ref{lcs}).

\subsection{System parameters}
\label{sec:sed_fit}

To obtain additional constraints on \s's system parameters, we modelled its observed broadband spectral energy distribution (SED) from near-UV to infrared (IR) wavelengths. We compiled the SED using archival photometry along with our OM observation. Specifically, we retrieved optical $v$, $g$, $r$, $i$, and $z$-band data from the SkyMapper Southern Survey catalogue~\citep[][]{Wolf2018},\footnote{\url{https://cdsarc.cds.unistra.fr/viz-bin/cat/II/358}} and $B$, $V$, as well as Sloan $g$, $r$, and $i$-band magnitudes from the AAVSO Photometric All Sky Survey~\citep[APASS,][]{Henden2015}.\footnote{\url{https://cdsarc.cds.unistra.fr/viz-bin/cat/II/336}} Additional optical data were taken from \textit{Gaia} DR3 $G$, $BP$, and $RP$ bands~\citep{Gaia2016, Gaia2023}\footnote{\url{https://cdsarc.cds.unistra.fr/viz-bin/cat/I/355}} and the TESS mean magnitude $Tr$~\citep{Paegert2021}.\footnote{\url{https://cdsarc.cds.unistra.fr/viz-bin/cat/IV/39}} Near-IR $J$, $H$, and $K$ data were obtained from the Two Micron All Sky Survey~\citep[2MASS,][]{Cutri2003,Skrutskie2006},\footnote{\url{https://cdsarc.cds.unistra.fr/viz-bin/cat/II/246}} and IR $W1$, $W2$, and $W3$ magnitudes from the All-Sky Wide-field Infrared Survey Explorer~\citep[AllWISE,][]{Wright2010}.\footnote{\url{https://cdsarc.cds.unistra.fr/viz-bin/cat/II/328}} For the UV regime, we used the mean UVM2-band flux/magnitude from our OM observation and archival UVW2-band data from \textit{Swift} Ultra-Violet/Optical Telescope~\citep[\textit{Swift}/UVOT,][]{Roming2005}. We note that \s\ was not included in the \textit{Swift}/UVOT point-source catalogue \citep{Yershov2014,Page2014},\footnote{\url{https://cdsarc.cds.unistra.fr/viz-bin/cat/II/339}} likely due to its high brightness causing saturation and pipeline failure. Nevertheless, given the importance of UV data to better constrain the mass accretion rate ($\dot{M}$), we manually extracted the UVW2 data and treat them as lower limits.

We modelled the observed SED of \s\ as the sum of four system components: the accretion disc, the white dwarf, a potential boundary layer, and the donor star. A detailed description of the modelling procedure and implementation of these components is provided in Appendix \ref{sed_appendix}. All parameter values adopted in the SED modelling are listed in Table \ref{sed_param}. From the fit, we obtained $\dot{M}$ and the system inclination ($i$), which were treated as free parameters. Using the derived $\dot{M}$ and the relation in Eq. \ref{eq:T_wd}, we inferred the white dwarf temperature ($T_{\rm WD}$). The best-fitting values, along with their 1$\sigma$ uncertainties, are given in the lower section of Table~\ref{sed_param}.

\begin{table}
\caption{System parameters adopted and derived for \s\ from SED modelling.}             
\label{sed_param}      
\centering        
\resizebox{\columnwidth}{!}{
\begin{tabular}{@{}l c c c}
\hline
\\[-1.5ex]
\multicolumn{4}{c}{\textit{Adopted for the fit}} \\
\\[-2ex]
\hline            
\rule{0pt}{2.5ex}Parameter &  & Value & Reference \\
\\[-2ex]
\hline     
   \rule{0pt}{2.3ex}Inner radius of the disc & $R_{\rm in}$ [$R_{\odot}$] & 0.0134 & (1) \\
   \rule{0pt}{2ex}Outer radius of the disc & $R_{\rm out}$ [$R_{\odot}$] & 0.531 & (2)\\
   \rule{0pt}{2ex}Mass of the white dwarf & $M_{\rm WD}$ [$M_{\odot}$] & 0.83 & (3)\\
   \rule{0pt}{2ex}Radius of the white dwarf & $R_{\rm WD}$ [$R_{\odot}$] & 0.0122 & (4)\\
   \rule{0pt}{2ex}Surface gravity of the white dwarf & $\log(g_{\rm WD})$ & 8.19 & (5)\\
   \rule{0pt}{2ex}Temperature of the boundary layer & $T_{\rm BL}$ [K] & $10^5$ & (6)\\
   \rule{0pt}{2ex}Width of the boundary layer & $\Delta R_{\rm BL}$ [$R_{\odot}$] & 0.00122 & (7)\\
   \rule{0pt}{2ex}Mass of the donor & $M_{\rm 2}$ [$M_{\odot}$] & 0.25 & (8)\\
   \rule{0pt}{2ex}Radius of the donor & $R_{\rm 2}$ [$R_{\odot}$] & 0.34 & (8)\\
   \rule{0pt}{2ex}Temperature of the donor & $T_{\rm 2}$ [K] & 3375 & (8)\\
   \rule{0pt}{2ex}Surface gravity of the donor & $\log(g_{\rm 2})$ & 4.78 & (8)\\
   \rule{0pt}{2ex}Semi-major axis & $A$ [$R_{\odot}$] & 1.22 & (9)\\
\hline
\\[-1.5ex]
\multicolumn{4}{c}{\textit{Derived from the fit}} \\
\\[-2ex]
\hline            
\rule{0pt}{2.5ex}Parameter &  & Value & Reference \\
\\[-2ex]
\hline            
    \rule{0pt}{2.3ex}Mass accretion rate & $\dot{M}$ [$10^{-9}$ M$_{\odot}$ yr$^{-1}$] & $6.8 \pm 0.7$ & (10)\\
    \rule{0pt}{2ex}Inclination & $i$ [\textdegree] & $56 \pm 2$ & (10)\\
    \rule{0pt}{2ex}Temperature of the white dwarf & $T_{\rm WD}$ [K] & $\sim$46000 & (10)\\
\hline
\end{tabular}} 
\tablebib{(1) \citet[][$1.1 \times R_{WD}$]{Frank2002}; (2) \citet[][$0.9 \times R_{L1}$]{Papaloizou1977}; (3) \citet{Zorotovic2011}; (4) \citet{Bedard2020}; (5) Based on $M_{WD}$ and $R_{WD}$; (6) \citet{Nabizadeh2020}; (7) $0.1 \times R_{WD}$, between $R_{WD}$ and $R_{in}$; (8) \citet[][$P_{\rm orb}$ = 3.6 h]{Knigge2011}; (9) Based on $M_{WD}$, $M_2$, and $P_{\rm orb}$; (10) This work.}
\end{table}

\begin{figure*}
 \includegraphics[trim=0mm 0mm 0mm 0mm,width=\textwidth]{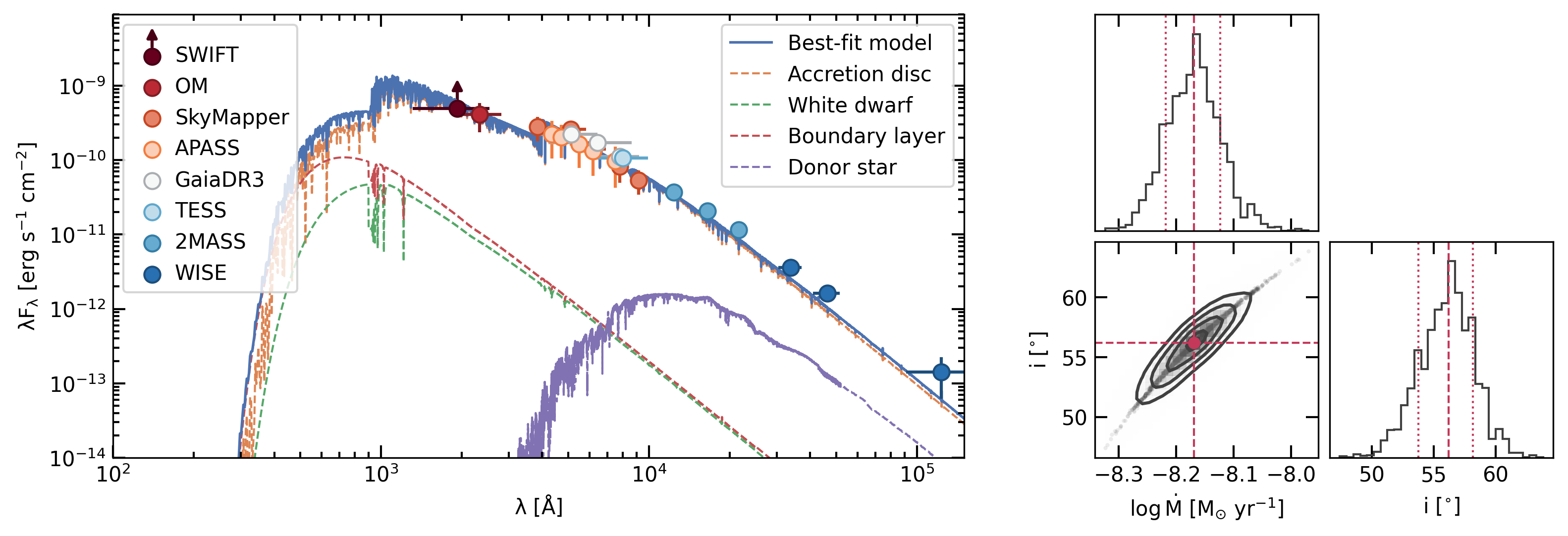}
 \caption{\textit{Left panel:} SED of \s\ as a function of wavelength, shown with the best-fitting model including contributions from the accretion disc, white dwarf, boundary layer, and donor star. \textit{Right panel:} Corner plot of the posterior distributions from the SED fit, showing the joint and marginal probability densities for the mass-accretion rate ($\dot{M}$) and system inclination ($i$). The diagonal panels display the 1D marginalised Bayesian posteriors; the red dashed lines mark the median values and the red dotted lines indicate the 68\% (1$\sigma$ for a Gaussian distribution) credible intervals. The off-diagonal panel shows the 2D joint contours at the nominal 68\%, 95\%, and 99.7\% (1$\sigma$, 2$\sigma$, and 3$\sigma$ for a Gaussian distribution) credible intervals, illustrating the covariance between $\log(\dot{M})$ and $i$.}
 \label{sed}
\end{figure*}

The left panel of Fig. \ref{sed} shows the resulting SED along with the best-fitting model and its individual components. The observed SED exhibits a steep blue continuum, similar to that seen in the optical spectra \citep[][]{Schwope2022}, consistent with a dominant contribution from the accretion disc. Beyond approximately $10^{4} \mathrm{\mathring{A}}$, however, an IR excess becomes apparent, as observed in other nova-like CVs \citep{Dhillon2000,Hoard2014}. This IR upturn has been attributed to circumbinary dust \citep{Hoard2009} or free-free emission from disc winds \citep{Dubus2003}. Additionally, there is a subtle trend for higher-inclination nova-like CVs to exhibit redder IR colours \citep{Hoard2002,Hoard2014}, possibly due to the increasingly edge-on view of the cooler outer disc and rim at larger inclinations. Overall, the SED fitting provides a robust estimate of $\dot M$ and $i$ for \s\ in its quasi-steady accretion state. However, we note a significant degeneracy between $i$ and $\dot{M}$, evident from the diagonal trend in the 2D covariance plot shown in the right panel of Fig. \ref{sed}.

\section{Discussion}

Although previous multi-wavelength studies left the subclassification of the CV \s\ uncertain (proposing it could be either an X-ray underluminous IP or an overluminous, non-magnetic nova-like, \citealt{Schwope2022}), this study aims to clarify its nature by exploring both hypotheses. We analysed the source's variability on short (minutes to hours) and long (months to years) timescales using optical, UV, and X-ray light curves, and investigated its X-ray spectral properties based on a dedicated 46 ks \xmm\ observation, complemented by the full \ero\ dataset and archival ASAS-SN optical photometry. In addition, we modelled the broadband SED, from near-UV to IR, to better constrain the system parameters.

\subsection{X-ray and UV periodicities}
\label{modulations}

\s\ displays significant variability on short timescales, as evidenced by the \xmm\ light curves in Fig. \ref{lcs} (panels 2-5), as well as on long timescales, as can be seen in the \ero\ light curve in the top panel of Fig. \ref{lcs}. The Lomb-Scargle power spectra of both the EPIC-pn and OM 30 s binned light curves revealed several modulations with a significance \mbox{>5$\sigma$} (see Fig. \ref{ls} and Table \ref{peaks}). In particular, a modulation at \mbox{$3.6\pm0.5$ h} is evident in both datasets, with a significance >8$\sigma$. This value is typical of the orbital periods of nova-like and IP CVs \citep[e.g.][]{Inight2023}, and it is fully consistent with a strong modulation \mbox{($3.645\pm0.006$ h)} found from optical spectroscopy by Brink et al. (in prep.). We therefore tentatively identify the 3.6 h modulation as the orbital period of \s\ and the 2 h and 1.4 h modulations as its harmonics (see Section \ref{timing}). 

Given its large error, our tentative orbital period is also consistent with the \mbox{$3.94\pm0.01$ h} modulation found in the first TESS dataset \citep{Pichardo-Marcano2020,Schwope2022}, which was subsequently confirmed by a 400-day interval ASAS-SN dataset \citep[$3.914\pm0.001$ h,][]{Schwope2022}. However, this 3.9 h modulation is inconsistent with the 3.645 h value reported by Brink et al. (in prep.), and it was absent in the second TESS dataset obtained two years later. Contrary to the initial association with $P_{\rm orb}$ proposed by \citet{Pichardo-Marcano2020}, Brink et al. (in prep.) interpret the TESS modulation as a likely positive superhump period, produced by periodic tidal stresses from the secondary acting on the disc, which becomes eccentric and subsequently precesses \citep{Whitehurst1988,Lubow1991}. 

As X-rays and UV emissions in CVs are expected to originate close to the white dwarf (either from its surface, an accretion column, or a boundary layer, \citealt{Patterson1985b,Mukai2017}), these spectral bands are well suited to detecting potential modulations at the white dwarf's spin period. Considering the possibility of a magnetic white dwarf in \s, our timing analysis in Section \ref{timing}, along with tentative correlations among several detected modulations, suggests two candidate spin periods: $36.0\pm0.7$ min and $43\pm1$ min. Both periods present meaningful correlations (see Table \ref{corr}), including possible $P_{\rm beat}$ ($(\omega-\Omega)^{-1}$ or $(\omega+\Omega)^{-1}$) and harmonics, with the other significant peaks found in the EPIC-pn and OM power spectra. However, none of these modulations were detected in the optical data of \s\ (Brink et al. in prep.). 

If \s\ were indeed an IP, the discrepancy between the modulations observed at different wavelengths would not be entirely unexpected, given the known variability in $P_{\rm spin}$ detection at different wavelengths in IPs. For example, in V1223 Sgr, $P_{\rm spin}$ is prominent in X-rays, while the dominant modulation in the optical is $P_{\rm beat}$ \citep[see][and references therein]{Jablonski1987}. Furthermore, the optical power spectra of FO Aqr, an IP exhibiting high and low optical states, presented by \citet[][see their fig. 13]{Littlefield2020}, revealed that $P_{\rm spin}$ dominates the high states while $P_{\rm beat}$ and its harmonics dominate the low states. Additionally, the optical, UV, and X-ray emissions from CVs are expected to originate from different regions within the system. While the disc does not reach temperatures high enough to emit X-rays, it is a source of IR, optical an UV radiation \citep{Mukai2017}. In contrast, the regions near the white dwarf, as previously discussed, are mostly responsible for producing X-rays, but also UV emissions. In the case of \s, we observe similar variability in both UV and X-rays. However, the UV light curve shows stronger modulation at the potential $P_{\rm orb}$, while the X-rays are more significantly modulated at the tentative $P_{\rm spin}$. This suggests that the emissions in these bands may originate from slightly different regions in the system.

Although the origin of each modulation detected in \s\ is not yet fully understood, our results are somewhat comparable with previous findings. In addition to the 3.9 h TESS modulation, \citet{Schwope2022} reported a \mbox{$2080\pm100$ s} ($35\pm2$ min) modulation from SAAO high-speed photometry, as well as a $\sim$25 min modulation from the second TESS observation. The former is fully consistent with our 36~min EPIC-pn modulation. The latter, while not directly comparable to any of the periodicities identified in this work, agrees with a 24.9 min modulation found by Brink et al. (in prep.) from new SAAO photometry. 

The presence of short-timescale, aperiodic variability (flickering) in the EPIC-pn light curve (Fig. \ref{lcs}), previously analysed in Section \ref{timing}, points to intrinsic variability in the source's emission rather than being caused by absorption effects. While the exact origin remains uncertain, such flickering may result from varying shocks in a potential boundary layer or from rapid fluctuations in the accretion flow onto the white dwarf, possibly occurring along an extended accretion arc.

\subsection{On the nature of the CV}
The ASAS-SN long-term optical light curve (see Fig. \ref{opt_lc}) reveals two distinct drops in brightness: one of $\sim$3 mag, followed by another of $\sim$1 mag. These declines are characteristic of \mbox{VY Scl-type} nova-likes and firmly establish \s\ as a nova-like CV. However, while nova-likes are typically considered non-magnetic, the timing properties discussed in the previous section suggest the possible presence of a magnetic white dwarf in this system.

From our spectral analysis of \s\ (see Section \ref{spec_ana}) we find that the best-fitting model to the \xmm\ spectra reveals a multi-temperature plasma, with components at \mbox{$kT_1 = 0.23$ keV}, \mbox{$kT_2 = 0.94$ keV}, and \mbox{$kT_3 = 5.2$ keV}. These temperatures likely correspond to emission from regions close to the white dwarf, such as a post-shock region, boundary layer, corona, or a combination of these. The plasma temperatures derived from the \ero\ spectra \mbox{($kT = 2.3$}, 13, 9, and \mbox{4.8 keV)} differ from the \xmm\ results but are broadly comparable once their large uncertainties are taken into account (see Table \ref{spec_param}). Non-magnetic CVs of the dwarf novae type typically exhibit plasma temperatures below \mbox{10 keV} \citep[][]{Byckling2010}, while IPs tend to show higher values in the \mbox{$\simeq$10$-$50 keV} range \citep[e.g.][]{Barlow2006,Landi2009,Brunschweiger2009}. The inferred temperatures in \s\ are therefore somewhat low for a typical IP and more consistent with non-magnetic CVs. However, we note that neither \xmm\ nor \ero\ is particularly sensitive to high-energy emission, although both remain useful for constraining high-temperature plasma components \citep[e.g.][]{Worpel2020}. For comparison, an \xmm\ analysis of the nova-like MV Lyr revealed plasma temperatures of \mbox{$kT_1$$\sim$1 keV} and \mbox{$kT_2$$\sim$6 keV} \citep{Dobrotka2017}, consistent with our findings and supportive of a nova-like classification for \s. Furthermore, our best-fitting model did not require a partial-covering absorber (commonly used in IP spectral models to account for X-ray emission from the accretion column, \citealt{Norton1989,Mukai2017}), which also favours a non-magnetic interpretation. Nonetheless, low plasma temperatures have been observed in the prototypical IP: DQ Her. In that case, an \xmm\ analysis yielded \mbox{$kT_{1} = 0.27$ keV} and \mbox{$kT_{2} = 0.84$ keV}, interpreted as evidence of extensive reprocessing \citep{Worpel2020}, although a partial-covering absorber was required. Similarly, the IP nature of XMMU J185330.7-012815 was confirmed based on \xmm\ data, modelled with a multi-temperature plasma (\mbox{$kT_1 = 0.65$ keV}, \mbox{$kT_2 = 1.78$ keV}, \mbox{$kT_3 = 10.84$ keV}), no partial-covering absorber, and a hydrogen column density of \mbox{$\sim5\times10^{20}$ cm$^{-2}$} \citep[][]{Hui2012}, a spectral configuration similar to that of \s. These cases suggest that, despite the relatively low plasma temperatures, our spectral analysis does not rule out a magnetic nature. Finally, while polars generally exhibit lower plasma temperatures than IPs \citep[e.g][]{Schwope2007,Schwope2020}, the presence of an accretion disc in \s---evidenced by its optical light curve (Fig. \ref{opt_lc}), SED (Section \ref{sec:sed_fit} and Fig. \ref{sed}), and the frequency break identified in the optical periodogram (Brink et al. in prep.)---effectively rules out a polar classification. 

Table \ref{spec_param} lists the unabsorbed bolometric (\mbox{0.01$-$100 keV}) luminosities for all epochs. Our values, \mbox{$0.63-5.3\times10^{32}$ erg s$^{-1}$}, are of the same order as the \mbox{$4.8\times10^{32}$ erg s$^{-1}$} bolometric luminosity reported by \citet{Schwope2022} for eRASS2; for that epoch we obtain \mbox{$(5.3\pm0.2)\times10^{32}$ erg s$^{-1}$}. The unabsorbed \mbox{0.25$-$10 keV} X-ray luminosities of \s\ (\mbox{$\gtrsim10^{32}$ erg s$^{-1}$}; Table \ref{spec_param}) place it between the most luminous non-magnetic CVs, whose luminosities in this band typically span $10^{30}-10^{32}$ erg s$^{-1}$ \citep[][see also \citealt{Byckling2010,Inight2022}]{Wada2017}, and IPs, which generally exhibit higher X-ray luminosities \citep[][]{deMartino2020,Schwope2024b}. Moreover, we note the existence of  low-luminosity IPs, identified by their optical brightness, which is strongly correlated with their low X-ray luminosities \citep[$L_{\rm X}<10^{32}$ erg s$^{-1}$,][]{Mukai2023}. Interestingly, a few members of the emerging population of magnetic nova-likes (discussed below) fall into this low-luminosity IP category. 

The Fe K$\alpha$ complex, including the fluorescent line at 6.4 keV produced by reflection, is observed across all types of CVs. Although a broad range of EW values has been reported in the literature, the average EW of the fluorescent line in dwarf novae is $\sim$60 eV, while in IPs it is $\sim$115 eV \citep[see][and references therein]{Xu2016}. The EW derived for \s\ in Section \ref{spec_ana} (\mbox{$150^{+60}_{-50}$ eV}) is consistent with the values typically observed in IPs. However, the amount of reflection depends on several factors intrinsic to the source, regardless of system type. These include the abundance of the reflector, the angle between the reflecting surface and the line of sight, the photon index of the spectrum of the X-ray emission source, and, to a lesser extent, the hydrogen column density \citep{Done1997,Ezuka1999,Ishida2009}. As a result, the EW alone provides limited diagnostic power regarding the nature of the system.

The broadband SED fitting of \s\ described in Section \ref{sec:sed_fit}, revealed a disc-dominated system. We estimated a mass accretion rate of \mbox{$\sim7\times10^{-9}$ M$_{\odot}$ yr$^{-1}$}, a value consistent with those typically observed in nova-likes systems \citep[e.g.][]{Hewitt2020}, but also within the range reported for IPs \citep[e.g.][]{deMartino2020}. In addition, we inferred a white dwarf temperature of \mbox{$T_{\rm WD}\simeq46000 $ K} from the estimated $\dot{M}$, which is in good agreement with values derived for other nova-likes such as DW UMa \citep[\mbox{$T_{\rm WD}\simeq50000$ K},][]{Araujo-Betancor2002}, MV Lyr \citep[\mbox{$T_{\rm WD}\simeq47000$ K},][]{Hoard2004}, and ASAS J0714$+$7004 \citep[\mbox{$T_{\rm WD}\simeq41000$ K},][]{Inight2022}. 

The link between nova-likes that exhibit low states and IPs has been explored in previous studies. \citet[][and references therein; see also \citealt{Duffy2024}]{Hameury2002} investigated the potential presence of magnetised white dwarfs in VY Scl-type nova-likes. The lack of dwarf-novae outburst cycles typically expected when the mass transfer rates decrease (leading to a drop in brightness) can be attributed to a magnetic field truncating the inner accretion disc. Such a magnetic field can be directly confirmed through the detection of pulsations on $P_{\rm spin}$ \citep[e.g.][]{Coughenour2022,Potter2024} and/or circular polarisation caused by cyclotron radiation \citep[e.g.][]{Lima2025}, both indicative of the white dwarf's rotation. The observational evidence of spin modulations in white dwarfs within nova-likes is growing, as an increasing number of IPs exhibit the low states typical of VY Scl systems. These include DW Cnc \citep{Rodriguez-Gil2004}, V1323 Her \citep{Andronov2014}, FO Aqr \citep{Littlefield2016}, J183221.56-162724.25 \citep{Beuermann2022}, DO Dra \citep{Hill2022}, V515 And, V1223 Sgr, RX J2133+5107, V1025 Cen \citep{Covington2022}, SRGe J194401.8+284452 \citep{Kolbin2024}, AO Psc \citep{Garnavich1988}, Swift J2113.5+5422 \citep{Halpern2018}, V1062 Tau \citep{Lipkin2004}, and the IP candidate Swift J0746.3-1608 \citep{Bernardini2019}. Moreover, circular polarisation has been detected in some nova-like CVs of the SW Sex type \citep{Rodriguez-Gil2001, Lima2021}, a subclass defined by their distinctive optical spectroscopic features. \citet{Rodriguez-Gil2001} even proposed that SW Sex systems could be IPs with particularly high mass accretion rates.  

In the case of \s, we previously noted the presence of X-ray modulations that could tentatively be associated with $P_{\rm spin}$, although no corresponding signal is detected in the optical data. However, $P_{\rm spin}$ is often more clearly detected in X-rays than in optical wavelengths in IPs (see Section \ref{modulations}), and a lack of correlation between modulations at different wavelengths has already been observed in nova-likes, such as the SW Sex systems \citep{Lima2021}. The strong variability seen in the X-ray light curves of \s\ (on timescales ranging from minutes to years) is also a common characteristic of IPs \citep[e.g.][]{Norton1996}. In addition to this general variability, we detected a < 20~days rebrightening during the optical low state (Fig. \ref{opt_lc}), which could correspond to a dwarf-nova-type outburst expected when the mass transfer rate decreases. This possibility does not contradict the magnetic white dwarf scenario proposed by \citet{Hameury2002}, since their framework predicts an outburst cycle rather than only one or two isolated events. On the other hand, no circular polarisation has been detected in \s\ (see Brink et al. in prep. for details). Detecting polarisation can be particularly challenging in low-inclination systems, where the polarised emission from the region near the white dwarf may be diluted by the bright accretion disc \citep{Lima2021}. Although our SED fitting in Section \ref{sec:sed_fit} suggests an inclination of $56$\textdegree, this value corresponds to the best fit and is strongly correlated with the mass accretion rate, leading to some degeneracy in the derived parameters (see Section \ref{sec:sed_fit}). In fact, Brink et al. (in prep.) argue for a low inclination based on the small amplitude of the radial velocities of the H$\beta$ and H$\gamma$ emission lines.

Despite the uncertainty surrounding the magnetic properties of \s, the growing evidence for magnetic fields in nova-like CVs suggests that current classification schemes may need to be revisited and incorporated into broader models of CV evolution. Expanding the sample of confirmed magnetic nova-likes through dedicated polarimetric observations and high-sensitivity X-ray timing studies will be essential to quantifying the prevalence and role of magnetic fields in these systems.

\section{Conclusions}

We have investigated the nature of the CV \s\ using both dedicated \xmm\ observations and \ero\ survey data, complemented by archival broadband photometry from the near-UV to the IR. The ASAS-SN long-term optical light curve, together with our broadband SED modelling, firmly establishes \s\ as a nova-like system. X-ray spectral analysis reveals emission from a multi-temperature plasma, while timing analysis suggests a tentative orbital period of 3.6 h. The power spectra show additional short-timescale (minutes to hours) variability which, combined with the pronounced long-term (months to years) X-ray variability seen in the \ero\ light curve and the X-ray luminosities of $\gtrsim10^{32}$ erg s$^{-1}$, also makes \s\ a candidate IP. Under this hypothetical scenario, the detected modulations at 36 min and 43 min are both plausible candidates for the white dwarf's spin period. While the detection of such a spin modulation is often sufficient to confirm a magnetic nature, the uncertainty surrounding all observed modulations (across multiple datasets and wavelengths) combined with the nova-like classification, means that the magnetic nature of \s\ remains uncertain. A consistent detection of a spin-related modulation and/or circular polarisation would be required to confirm the IP scenario. Notably, while nova-likes are typically considered non-magnetic, a dual classification as both nova-like and IP is possible, as evidenced by the growing number of confirmed IPs that exhibit the typical low states of VY Scl nova-likes.

\begin{acknowledgements}
    We thank the anonymous referee for their useful and thoughtful comments, which helped to improve this paper. VAC acknowledges support from the Deutsches Zentrum für Luft- und Raumfahrt (DLR) under contract 50 OR 2405, and JK from DLR under contract 50 OR 2408. JB is supported by the Deutsche Forschungsgemeinschaft (DFG, German Research Foundation) through grant Schw536/37-1. MV acknowledges support from the Science and Technology Facilities Council (STFC) studentship ST/W507428/1. This research has made use of data and/or software provided by the High Energy Astrophysics Science Archive Research Center (HEASARC), which is a service of the Astrophysics Science Division at NASA/GSFC. Based on observations obtained with XMM-Newton, an ESA science mission with instruments and contributions directly funded by ESA Member States and NASA. This work is based on data from eROSITA, the soft X-ray instrument aboard SRG, a joint Russian-German science mission supported by the Russian Space Agency (Roskosmos), in the interests of the Russian Academy of Sciences represented by its Space Research Institute (IKI), and the Deutsches Zentrum für Luft- und Raumfahrt (DLR). The SRG spacecraft was built by Lavochkin Association (NPOL) and its subcontractors, and is operated by NPOL with support from the Max Planck Institute for Extraterrestrial Physics (MPE). The development and construction of the eROSITA X-ray instrument was led by MPE, with contributions from the Dr. Karl Remeis Observatory Bamberg \& ECAP (FAU Erlangen-Nuernberg), the University of Hamburg Observatory, the Leibniz Institute for Astrophysics Potsdam (AIP), and the Institute for Astronomy and Astrophysics of the University of Tübingen, with the support of DLR and the Max Planck Society. The Argelander Institute for Astronomy of the University of Bonn and the Ludwig Maximilians Universität Munich also participated in the science preparation for eROSITA. The eROSITA data shown here were processed using the eSASS software system developed by the German eROSITA consortium. This work incorporated data provided by ASAS-SN. ASAS-SN is funded by Gordon and Betty Moore Foundation grants GBMF5490 and GBMF10501 and the Alfred P. Sloan Foundation grant G-2021-14192. This work made use of \texttt{astropy} (\url{http://www.astropy.org}): a community-developed core Python package and an ecosystem of tools and resources for astronomy \citep{Robitaille2013,Price-Whelan2018,Astropy2022}. We also made use of \texttt{numpy} \citep{Harris2020}, \texttt{scipy} \citep{Virtanen2020}, and \texttt{matplotlib} \citep{Hunter2007} \textsc{python} packages.
\end{acknowledgements}

\bibliographystyle{aa} 
\bibliography{references}

\begin{appendix}

    \section{Phase-folded light curves}
    
\begin{figure*}[!ht]
     \includegraphics[trim=0mm 0mm 0mm 0mm,width=\textwidth]{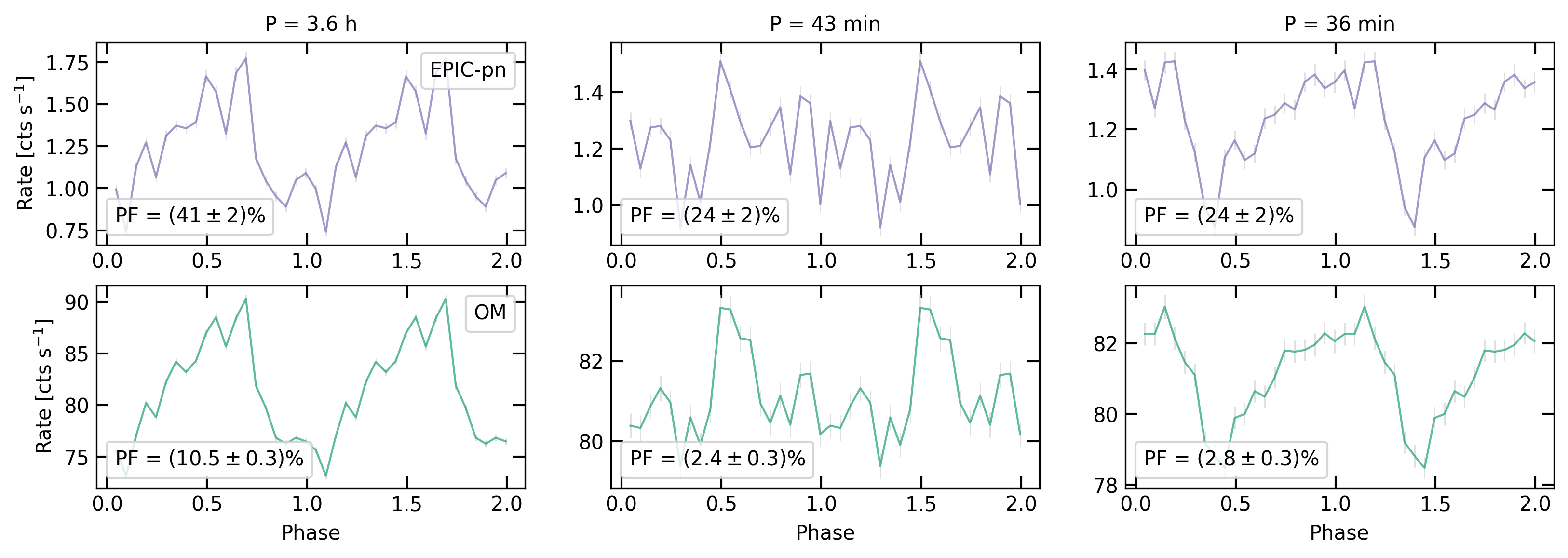}
     \caption{\xmm/EPIC-pn (top panels) and OM (bottom panels) light curves folded on the 3.6 h, 43 min, and 36 min modulations. Two cycles are shown for clarity. In all cases, phase zero is defined arbitrarily at the start of the EPIC-pn dataset (MJD 60055.498855949016). PF denotes the pulse fraction.}
     \label{foldedLCs}
    \end{figure*}

    Fig. \ref{foldedLCs} shows the \xmm/EPIC-pn and OM light curves folded on the 3.6 h, 43 min, and 36 min modulations discussed in Section \ref{timing}. The data were folded using 20 phase bins, with phase zero set arbitrarily at the start of the EPIC-pn dataset (MJD 60055.498855949016). The pulse fraction (PF), which quantifies the degree of modulation, was estimated as

    \begin{equation}
        PF = \frac{R_{max}-R_{min}}{R_{max}+R_{min}},
    \end{equation}

    \noindent where $R_{max}$ and $R_{min}$ are the maximum and minimum count rates in the folded light curve, respectively.

    \section{SED modelling}
    \label{sed_appendix}

    We modelled the broadband spectral energy distribution (SED) of \s\ (Section \ref{sec:sed_fit}), accounting for four key components of the system:

    \begin{enumerate}
    \item[-]\textbf{Accretion disc.} In nova-like CVs, the accretion disc is the primary source of near-UV to IR emission and is expected to dominate the observed SED \citep{Belyakov2010,Inight2022}. We modelled the disc as a set of $N = 1000$ concentric annuli extending from the inner radius ($R_{\rm in}$) to the outer radius ($R_{\rm out}$). The local effective temperature ($T_{\rm eff}$) of each annulus was determined by the flux per unit surface area ($Q$) at a given radius ($R$) in the disc \citep[][see equation 2.6 in the latter]{Lynden-Bell1969,Shakura1973}:

    \begin{equation}
    Q (R) = \frac{3 G M_{\rm WD} \dot{M}}{8 \pi R^3}\left(1 - \sqrt{\frac{R_{\rm in}}{R}}\right) = \sigma T_{\rm eff} (R)^{4}.
    \label{eq:T_disc}
    \end{equation}
    
Here $M_{\rm WD}$ and $\dot{M}$ are the white dwarf mass and the mass accretion rate, respectively. The inner radius was assumed to be $R_{\rm in} = 1.1~ R_{\rm WD}$ \citep{Cannizzo1993,Frank2002}, where $R_{\rm WD}$ is the radius of the white dwarf. $R_{\rm out}$ was set to 90\% of the white dwarf's Roche lobe radius ($R_{\rm L1}$), following \citet{Papaloizou1977} and \citet{Paczynski1977}. $R_{\rm L1}$ can be approximated in terms of the system's semi-major axis ($A$) as \citep[][equation 2]{Eggleton1983}
    
    \begin{equation}
    R_{\rm out} = 0.9 \times R_{\rm L1} = 0.9 \times \frac{0.49 q^{2/3}}{0.6 q^{2/3} + \ln(1 + q^{1/3})}\times A\,,
    \label{eq:roche}
    \end{equation}
    
    where $q = M_{\rm WD}/M_{\rm 2}$ is the white dwarf-to-donor star mass ratio.

    We interpolated the spectrum of each disc annulus using stellar atmosphere models from the BT-Settl grid \citep{Allard2012}\footnote{\url{https://svo2.cab.inta-csic.es/theory/newov2/index.php?models=bt-settl-agss}} at the corresponding $T_{\rm eff}(R)$ (Eq. \ref{eq:T_disc}), assuming a constant surface gravity $\log(g) = 4.0$ and solar metallicity. In reality, the surface gravity in the disc varies with radius, following $\log(g) = \Omega_{\rm K}^{2}\,z$, where $\Omega_{\rm K}$ is the Keplerian angular velocity and $z$ is the height from the disc mid-plane. This results in $\log(g)$ values ranging from $\sim$6--6.5 in the inner regions down to $\sim$3 at the outer edge. However, in the absence of synthetic spectra computed for such extreme $\log(g)$ conditions, we adopted a representative median value of $\log(g) = 4.0$. While the limitations of applying synthetic stellar spectra to model accretion disc emission are well known \citep{Wade1988,Hubeny2021}, this approach nonetheless offers a more realistic alternative to the overly simplistic blackbody approximation. 

    \item[-]\textbf{White dwarf.} The white dwarf temperature ($T_{\rm WD}$) in CVs has been extensively studied by \citet{Townsley2009}, who showed that ongoing accretion leads to significant heating of the white dwarf's surface. Their analysis demonstrated that $T_{\rm WD}$ typically ranges from $\sim$10000 K to $\sim$50000 K, depending primarily on the system’s long-term average mass accretion rate $\langle \dot{M} \rangle$. They also provided a scaling relation quantifying this dependence as
    
    \begin{equation}
    T_{\rm WD} = 1.7\times 10^4\ {\rm K} \, \left(\frac{\langle \dot{M} \rangle}{10^{-10}M_\odot\ {\rm yr^{-1}}}\right)^{1/4}\left(\frac{M_{\rm WD}}{0.9M_\odot}\right)\ ,
    \label{eq:T_wd}
    \end{equation}
    with $\langle \dot{M} \rangle \approx \dot{M}$ since we are dealing with a nova-like CV. 

    The white dwarf flux at the given temperature (Eq. \ref{eq:T_wd}) was modelled using a grid of synthetic spectra for hot white dwarfs \citep{Levenhagen2017}\footnote{\url{https://svo2.cab.inta-csic.es/theory/newov2/index.php?models=levenhagen17}}. The corresponding surface gravity was calculated based on an assumed white dwarf mass \mbox{$M_{\rm WD} = 0.83 \,M_{\odot}$} (the average value suggested for CVs by \citealt[][]{Zorotovic2011}) and a radius \mbox{$R_{WD}=0.0122\,R_{\odot}$}, obtained from the mass-radius relations of \citet[][]{Bedard2020}.\footnote{\url{https://www.astro.umontreal.ca/~bergeron/CoolingModels/}} We fixed $M_{\rm WD}$ in the SED modelling to avoid its strong degeneracy with the accretion rate $\dot{M}$ (Eq. \ref{eq:T_disc}).

    \item[-]\textbf{Boundary layer.} In CVs, a hot transition region known as the boundary layer is thought to exist where the accretion disc meets the surface of the white dwarf \citep[e.g.][]{Warner2003}. In this region, a substantial fraction of the gravitational potential energy is converted into heat as the Keplerian flow of the disc material slows abruptly to match the white dwarf's rotational velocity. Temperatures can reach $10^{5}$\,K in the optically thick part of the boundary layer, and $\sim$$10^{7-8}$\,K in the optically thin regions believed to emit X-rays \citep[][]{Jensen1984,Popham1995,Nabizadeh2020}. However, the boundary layer's detailed properties, including its geometry, energy-release efficiency, temperature profile, and even its very existence, remain uncertain, in part due to the strong attenuation of extreme-UV emission by interstellar extinction \citep[e.g.][]{Long1996,Hertfelder2013,Coleman2022}. 
    
    Since the boundary layer's emission peaks in the far/extreme UV (optically thick material), its contribution to the observed SED is expected to be minimal. To represent this component, we modelled a hot ring located at $R\approx R_{WD}$, with a width of $\Delta R_{\rm BL} =0.1 \times R_{\rm WD}$, positioned between the inner edge of the accretion disc and the white dwarf's surface. A temperature of $T_{\rm BL}=10^5$\,K was adopted, and the ring's emission was modelled using the same synthetic spectral grid used for hot white dwarfs \citep{Levenhagen2017}. The surface gravity was set to $\log(g) = 7$, the lowest available in the grid, and within the expected range ($\log(g)\approx6-8$) for this region.  

    \item[-]\textbf{Donor star.} The physical parameters of the donor star (mass, radius, surface gravity, and effective temperature) were adopted from the evolutionary models for CVs presented by \citet[][see their table~2]{Knigge2011},\footnote{\url{https://cdsarc.cds.unistra.fr/viz-bin/cat/J/ApJS/194/28}} based on the tentative orbital period of $3.6$\,h. Assuming solar metallicity, we interpolated the donor's spectrum from the same BT-Settl grid used for the accretion disc \citep{Allard2012}. 
    \end{enumerate}

    The interstellar absorption toward \s\ is relatively low, with $A_{\rm V} \lesssim 0.05$ according to \citet{Doroshenko2024},\footnote{\url{http://astro.uni-tuebingen.de/nh3d/nhtool}} consistent with our estimates of $N_{\rm H}$~(see Table~\ref{spec_param}). We applied extinction corrections using the \texttt{extinction} \texttt{Python} package \citep[][]{Barbary2021},\footnote{\url{https://extinction.readthedocs.io/en/latest/}} adopting $A_{\rm V} = 0.05 $ and the extinction law from \citet{Fitzpatrick1999}.

    We fitted the mass accretion rate ($\dot{M}$) and system inclination ($i$) by imposing a lower bound on the $\dot{M}$ prior to ensure the disc remains in a hot, stable state; specifically, requiring $T_{\rm eff} (R_{\rm out}) >= 6000$\,K \citep[current estimations lie within $6000-8000$\,K; see][and references therein]{Lasota2001}. A uniform prior was adopted for the inclination, ranging from 0\textdegree\ to 90\textdegree. The fit was performed within a Bayesian inference framework using the nested sampling algorithms \citep{Skilling2004,Skilling2006}, as implemented in the \texttt{dynesty} \texttt{Python} package \citep{Speagle2020}.\footnote{\url{https://dynesty.readthedocs.io/en/v2.1.5/}} This technique, although designed for computing Bayesian evidence, employs MCMC-style constrained sampling to explore the posterior, making it conceptually similar to conventional MCMC methods. To improve computational efficiency and avoid interpolation artefacts when using synthetic atmosphere models, we did not generate a full model spectrum for each point in parameter space. Instead, we precomputed a grid of synthetic photometric fluxes (e.g. $V, B, J$ bands) for each spectral model using the \texttt{pyphot} tool \citep{Fouesneau2025}.\footnote{\url{https://mfouesneau.github.io/pyphot/index.html}} During fitting, we interpolated the corresponding band fluxes from this precomputed grid using the \texttt{isochrones} interface \citep{Morton2015}.\footnote{\url{https://github.com/timothydmorton/isochrones}} The resulting synthetic photometry was then directly compared to the observed photometry. To account for both intrinsic variability of \s\ across different epochs and potential underestimation of measurement uncertainties, we followed the approach of \citet{Vines2022},\footnote{\url{https://github.com/jvines/astroARIADNE}} by introducing an additive excess noise term into the Gaussian log-likelihood for each band. This term was treated as a free parameter and allowed the photometric uncertainties to scale up to 10\% of the measured flux values. 

\end{appendix}

\label{LastPage}
\end{document}